\documentclass[aps,pre,twocolumn,floats]{revtex4}
\usepackage{graphicx}
\def\leqa{\stackrel{<}{\scriptstyle \sim}}

\begin{document}
%\noindent{\Large{\bf Ray dynamics in the AET experiment}}\\
\title{Ray dynamics in a long-range acoustic propagation experiment}

\author{Francisco J. Beron-Vera,$^{1}$ Michael G. Brown,$^{1}$
John A. Colosi,$^{2}$ Steven Tomsovic,$^{3}$ Anatoly L. Virovlyansky, $^{4}$
Michael A. Wolfson,$^{5}$ George M. Zaslavsky $^{6}$}

 \date{August 2002}

\affiliation{
$^1$ \mbox{Rosenstiel School of Marine and Atmospheric Science, University of Miami, Miami, Florida.} \\
$^2$ \mbox{Woods Hole Oceanographic Institution, Woods Hole, Massachusetts.} \\
$^3$ \mbox{Department of Physics, Washington State University, Pullman, Washington.} \\
$^4$ \mbox {Institute of Applied Physics, Russian Academy of Science, 6003600 Nizhny Novgorod, Russia.} \\
$^5$ \mbox {Applied Physics Laboratory, University of Washington, Seattle, Washington.} \\
$^6$ \mbox {Courant Institute of Mathematical Sciences, New York University, New York, New York.} \\
}

\begin{abstract}
A ray-based wavefield description is employed in the interpretation of
measurements
made during the November 1994 Acoustic Engineering Test (the AET experiment).
In this experiment phase-coded pulse-like signals with 75 Hz center frequency
and 37.5 Hz bandwidth were transmitted near the sound channel axis in the
eastern North Pacific Ocean.  The resulting acoustic signals were recorded on
a moored vertical receiving array at a range of 3252 km.  In our analysis both
mesoscale and internal-wave-induced sound speed perturbations are taken into
account.  Much of this analysis exploits results that relate to the subject of
ray chaos; these results follow from the Hamiltonian structure of the ray
equations.  We present evidence that all of the important features of the
measured AET wavefields, including their stability, are consistent with a
ray-based wavefield description in which ray trajectories are predominantly
chaotic.  
\end{abstract}

\maketitle

%\pacs{PACS numbers:   }

\section{Introduction}
The chaotic dynamics of ray trajectories in ocean acoustics have been explored
in a number of recent publications (\cite{AZ85}-\cite{rdoa}), including two
recent reviews~\cite{SVZ,rdoa}.  In these publications a variety of theoretical
results are presented and illustrated, mostly using idealized models of the
ocean sound channel.  In contrast, in the present study a ray-based wavefield
description is employed, in conjunction with a realistic environmental model,
to interpret a set of underwater acoustic measurements.  For this purpose we
use both oceanographic and acoustic measurements collected during the November
1994 Acoustic Engineering Test (the AET experiment).  In this experiment
phase-coded pulse-like signals with 75 Hz center frequency and 37.5 Hz
bandwidth were transmitted near the sound channel axis in the eastern North
Pacific Ocean.  The resulting acoustic signals were recorded on a moored
vertical receiving array at a range of 3252 km.  We present evidence that all
of the important features of the measured AET wavefields are consistent with
a ray-based wavefield description in which ray trajectories are predominantly
chaotic. 

The AET measurements have previously been analyzed in Refs.
\cite{AET1}-\cite{CDT}.  In the analysis of Worcester et al. \cite{AET1} it
was shown that the early AET arrivals could be temporally resolved, were
stable over the duration of the experiment and could be identified with
specific ray paths.  These features of the AET observations are consistent
with those of other long-range data sets \cite{SSM,S89a}.  The AET arrivals
in the last 1 to 2 sec (the arrival finale) could not be temporally resolved
or identified with specific ray paths, however.  Also, significant vertical
scattering of acoustic energy was observed in the arrival finale.  These
features of the AET arrival finale are consistent with measurements and
simulations (both ray- and parabolic-equation-based) at 250 Hz and 1000 km
range \cite{S89a}-\cite{CFB} and at 133 Hz and 3700 km range \cite{WS}. 

The analysis of Colosi et al. \cite{AET2} focused on statistical properties of
the early resolved AET arrivals.  Measurements of time spread, travel time
variance and probability density functions (PDFs) of peak intensity were
presented and compared to theoretical predictions based on a path integral
formulation as described in Ref. \cite{FDMWZ}.  Travel time variance was well
predicted by the theory, but time spreads were two to three orders of magnitude
smaller than theoretical predictions, and peak intensity PDFs were close to
lognormal, in marked contrast to the predicted exponential PDF.  The surprising
conclusion of this analysis was that the measured AET pulse statistics
suggested propagation in or near the unsaturated regime (characterized
physically by the absence of micromultipaths and mathematically by use of a
perturbation analysis based on the Rytov approximation), while the theory
predicted propagation in the saturated regime (characterized by the presence of
a large number of micromultipaths whose phases are random).

In Ref. \cite{CDT} it was shown that in the finale region of the measured AET
wavefields, where no timefronts are resolved, the intensity PDF is close to
the fully saturated exponential distribution.  This result is not unexpected
because the finale region is characterized by strong scattering of both rays
\cite{SFW} and modes \cite{CF}.
 
The work reported here seeks to elucidate both the physics underlying the AET
measurements, and the causes of the successes
and failures of the analyses reported in Refs. \cite{AET1}-\cite{CDT}.  We show
that a ray-based analysis, in which ideas associated with ray chaos play an
important role, can account for the stability of the early arrivals, their
small time spreads, the associated near-lognormal PDF for intensity peaks, the
vertical scattering of acoustic energy in the reception finale, and the
near-exponential intensity PDF in the reception finale.  Part of the success of
this interpretation comes about due to the presence of a mixture of stable and
unstable ray trajectories.  Also, an explanation is given for differing
intensity statistics in the early and late portions of the measured wavefields,
and the fairly rapid transition between these regimes.  

The remainder of this paper is organized as follows.  In the next section we
describe the AET environment and the most important features of the measured
and simulated wavefields.  In the simulated wavefields both measured mesoscale
and simulated internal-wave-induced sound speed perturbations are taken into
account.  It is shown that even in the absence of internal waves, the late
arrivals are not resolved and the associated ray paths
are chaotic. In section II we present results that relate to the
structure of the early portion of the measured timefront and its stability.
The micromultipathing process is discussed, and a quantitative explanation
for the remarkably small time spreads of the early ray arrivals is given.
Section III is concerned with intensity statistics.  An explanation is given
for the cause of the different wavefield intensity statistics in the early
and late portions of the arrival pattern.  In the final section we summarize
and discuss our results.
 
\section{ Measured and simulated wavefields}

Figure 1 shows a measured AET wavefield in the time-depth plane and ray-based
simulations of such a wavefield with and without internal waves.  Figure 2 is
a complementary display that shows measured and simulated wavefields after
plane wave beamforming.  In this section we describe the most important
qualitative features of measured and simulated AET wavefields.  We begin by
reviewing fundamental ray theoretical results so that our ray simulations can
be fully understood.  Also, this material provides a foundation for some of
the later discussion.  We then briefly describe the AET environment
and the signal processing steps that were performed to produce the measured
wavefield shown in Fig. 1.  Next, our treatment of internal-wave-induced sound
speed perturbations is briefly described.  Finally, we return to describing
and comparing qualitative features of the measured and simulated wavefields. 

Fixed-frequency (cw) acoustic wavefields satisfy the Helmholtz equation,
\begin{equation}
\label{Helm}
\nabla^2 u + \sigma^2 c^{-2}(z, r) u = 0,
\end{equation}
where $\sigma = 2 \pi f$ is the angular frequency of the wavefield and
$c(z, r)$ is the sound speed.  (The extension to transient wavefields
is straightforward using Fourier synthesis.  It is not necessary to consider
this complication to present the most important results needed below.)  We
assume propagation in the vertical plane with cartesian coordinates $z$ and
$r$ representing depth and range, respectively.  Ray methods can be introduced
when the wavelength
$2 \pi c/\sigma$ of all waves of interest is small compared to the length
scales that characterize variations in $c$.  Substituting the geometric ansatz,
\begin{equation}
\label{geom}
u(z, r;\sigma) = \sum_j A_j(z, r) e^{i \sigma T_j(z, r)},
\end{equation}
representing a sum of locally plane waves, into the Helmholtz equation gives,
after collecting terms in descending powers of $\sigma$, the eikonal equation,
\begin{equation}
\label{eik}
(\nabla T)^2 = c^{-2},
\end{equation}
and the transport equation,
\begin{equation}
\label{trans}
\nabla(A^2 \nabla T) = 0.
\end{equation}
For notational simplicity, the subscripts on $A$ and $T$ have been dropped in
(\ref{eik}) and (\ref{trans}).

The solution to the eikonal equation (\ref{eik}) can be reduced to solving a
set of ray equations.  If a one-way formulation (see, e.g., Ref. \cite{rdoa})
is invoked, with $r$ playing the role of the independent variable, the ray
equations are
\begin{equation}
\label{ray1}
\frac{dz}{dr} =  \frac{\partial H}{\partial p}, \;
\frac{dp}{dr} = -\frac{\partial H}{\partial z},
\end{equation}
and
\begin{equation}
\label{dTdr}
\frac{dT}{dr} = L = p \frac{dz}{dr} - H,
\end{equation}
where
\begin{equation}
\label{H1}
H(p,z,r) = - \sqrt{c^{-2}(z,r) - p^2}.
\end{equation}
These equations constitute a nonautonomous Hamiltonian system with one degree
of freedom; $z$ and $p$ are canonically conjugate position and momentum
variables, $r$ is the time-like variable, $H$ is the Hamiltonian and the
travel time $T$ corresponds to the classical action.  Note also that
$p \equiv p_z = \partial T/\partial z$ and
$p_r = \partial T/\partial r = -H$ are the $z$- and $r$-components,
respectively, of the ray slowness vector.  The ray angle relative to the
horizontal $\varphi$ satisfies $dz/dr = \tan \varphi$, or, equivalently,
$cp = \sin \varphi$.  For a point source rays can be labeled by their
initial slowness $p_0$.  The solution to the ray equations is then $z(r;p_0)$,
$p(r;p_0)$, $T(r;p_0)$.  The terms in (\ref{geom}) correspond to eigenrays;
these satisfy $z(r_R; p_0) = z_R$ where the depth and range of the receiver
are $z_R$ and $r_R$, respectively.  The transport equation (\ref{trans}) can
be reduced to a statement of constancy of energy flux in ray tubes.  The
solution, for the $j$-th eigenray, can be written
\begin{equation}
\label{gampl}
A_j(z,r) = A_0 \: |H/q_{21}|_j^{1/2} \: e^{-i \mu_j \frac{\pi}{2}}
\end{equation}
where $H$, the matrix element $q_{21}$ and the Maslov index $\mu$ are
evaluated at the endpoint of the ray, and $A_0$ is a constant.  The stability
matrix
\begin{equation}
\label{Q}
Q =
\left (
\begin{array}{cc}
q_{11} & q_{12}\\ q_{21} & q_{22}
\end{array}
\right )
= \left (
\begin{array}{cc}
\left. \frac{\partial p}{\partial p_0} \right |_{z_0} &
\left. \frac{\partial p}{\partial z_0} \right |_{p_0}\\
\left. \frac{\partial z}{\partial p_0} \right |_{z_0} &
\left. \frac{\partial z}{\partial z_0} \right |_{p_0}
\end{array}
\right ) 
\end{equation}
quantifies ray spreading,
\begin{equation}
\label{vareq}
\left (
\begin{array}{c}
\delta p \\
\delta z
\end{array}
\right )
= Q
\left (
\begin{array}{c}
\delta p_0 \\
\delta z_0
\end{array}
\right ) .
\end{equation}
Elements of this matrix evolve according to
\begin{equation}
\label{dQdr}
\frac{d}{dr}Q = KQ
\end{equation}
where $Q$ at $r = 0$ is the identity matrix, and
\begin{equation}
\label{K}
K =
\left (
\begin{array}{cc}
-{\partial ^2 H\over \partial z \partial p} &
-{\partial ^2 H\over \partial z^2} \\
{\partial ^2 H\over \partial p^2} &
{\partial ^2 H\over \partial z \partial p}
\end{array}
\right ).
\end{equation}
At caustics $q_{21}$ vanishes and, for waves propagating in two space
dimensions, the Maslov index advances by one unit.

In ocean environments with realistic range-dependence, including the AET
environment, ray trajectories are predominantly chaotic.  Chaotic rays
diverge from neighboring rays at an exponential rate, on average, and are
characterized by a positive Lyapunov exponent,
\begin{equation}
\label{Lyap}
\nu_L = \begin{array}{c} \lim \\ r\!\rightarrow\!\infty \end{array} \left(
  \frac{1}{r} \begin{array}{c} \lim \\ {\cal D}(0)\!\rightarrow\!0
\end{array}
       \ln \frac{{\cal D}(r)}{{\cal D}(0)} \right).
\end{equation}
Here ${\cal D}(r)$ is a measure of the separation between rays at range $r$;
suitable choices for ${\cal D}$ are any of the four matrix elements of $Q$
or the trace of $Q$.  The chaotic motion of ray trajectories leads to some
limitations on predictability.  This will be discussed in more detail below.
Additional background material relating to this topic can be found in
\cite{SVZ,rdoa} and references therein.

We digress now to briefly describe the AET experiment and relevant signal
processing.  In this experiment a source, suspended at 652 m depth from the
floating instrument platform R/P FLIP in deep water west of San Diego,
transmitted sequences of a phase-coded signal whose autocorrelation is
pulse-like (with a pulse width of approximately 27 ms).  The source center
frequency was 75 Hz and the bandwidth was approximately 37.5 Hz.  The resulting
acoustic signals were recorded east of Hawaii at a range of 3252.38 km on a
moored 20 element receiving array between the depths of 900 m and 1600 m.
After correcting for the motion of the source and receiving array, including
removal of associated Doppler shifts, the received acoustic signals were
cross-correlated with a replica of the transmitted signal (to achieve the
desired pulse compression) and complex demodulated.  To improve the
signal-to-noise ratio, 28 consecutive processed receptions, extending over a
total duration of 12.7 minutes, were coherently added.  The duration over which
this coherent processing yields improved signal-to-noise is limited by ocean
fluctuations, mostly due to internal waves.  The statistics of the acoustic
fluctuations were shown in Ref. \cite{CDT} not to be adversely affected by the
12.7 minute coherent integration.  Nearly concurrent with the week-long
experiment, temperature profiles in the upper 700 m were measured with XBT's
every 30 km along the transmission path.
An objective mapping technique was applied \cite{AET1} to these measurements
to produce a map of the background sound speed structure (including mesoscale
structure) along the transmission path.  The sampling interval in
range of the resulting sound speed field is 30 km so that no structure with
horizontal wavelengths less than 60 km is present in this field.  More details
on the AET experiment, environment and processing of the acoustic signals can
be found in Ref. \cite{AET1}.

Internal-wave-induced sound speed perturbations are taken into account in
most of our simulations.  These are assumed to satisfy
the relationship $\delta c = (\partial c/\partial z)_p \zeta$ and the
statistics of $\zeta$, the internal-wave-induced vertical displacement of a
water parcel, were assumed to be described by the empirical Garrett-Munk
spectrum (see, e.g., Ref. \cite{Miw}).  The potential sound speed gradient
$(\partial c/\partial z)_p$, the buoyancy frequency $N$, and the sound speed
$c$ were estimated directly from hydrographic measurements using a procedure
similar to that described in the appendix of Ref. \cite{WS}; it was found that
a good approximation for the AET environment is
$(\partial c/\partial z)_p/c \approx (1.25 \; {\rm s}^2/{\rm m}) N^2$.
The vertical displacement $\zeta(r,z)$ was computed using equation 19 of
Ref. \cite{CB} with the variable $x$ replaced by $r$, and $y=t=0$.  Physically
this corresponds to a frozen vertical slice of the internal wave field that
includes the influence of transversely propagating internal wave modes.  A
Fourier method, with $\Delta k_x = 2\pi/3276.8$ km and $0 < k_x \leq 2\pi/1$
km, was used to numerically generate sound speed perturbation fields.  (Ray
calculations were also performed using an internal wave field in which the hard
upper bound on $k_x$ was replaced by an exponential damping of the large $k_x$
energy.  The qualitative ray behavior described below did not depend on the
manner in which the large $k_x$ cut-off was treated.)  A mode number cut-off
of $j_{max} = 30$ was used in all the simulations shown.  Contributions from
approximately $2^{30}$ ($2^{15}/5 \times 2^{15}/5 \times 30$ where the factors
of $1/5$ are present because the assumed wavenumber cut-off was one-fifth of
the assumed Nyquist wavenumber) internal wave modes were included in our
simulated internal-wave-induced sound speed perturbation fields.  Measurements
of sound speed variance made during the experiment (see Fig. 3) indicate that
the internal wave energy strength parameter $E$ was close to $E_{GM}$, the
nominal Garrett-Munk value.  Simulations were performed using both
$E = 0.5 \: E_{GM}$ and $E = 1.0 \: E_{GM}$.  All of the simulations shown in
this paper use $E = 0.5 \: E_{GM}$.  The difference between these simulations
and those performed using $E = 1.0 \: E_{GM}$ will be discussed where
appropriate. 

With the foregoing discussion as background, we are prepared to discuss our
ray simulations in the AET environment.  These are shown in Figs. 1, 2, 4
through 11, and 14.  The points plotted in Figs. 1, 2 and 4 were computed by
integrating the coupled ray equations (\ref{ray1}, \ref{dTdr}) from the source
to the range of the receiving array.  Several fixed and variable step-size
integration algorithms were tested.  In the presence of internal waves,
convergence using a fixed step fourth order Runge Kutta integrator required
that the step size not exceed approximately one meter.  All ray tracing
calculations presented in this paper were performed using double precision
arithmetic (64 bit floating point wordsize).  The wavefield intensity is
approximately proportional to the density of rays (dots) that are plotted in
Figs. 1 and 2.  A somewhat more difficult calculation in which the
contributions to the wavefield from many rays are coherently added will be
discussed below. 

In the presence of internal waves, rays with launch angles between
approximately $\pm5^o$ form the diffuse finale region of the arrival pattern
arriving after approximately 2196 sec.  Steeper rays contribute to the earlier,
mostly resolved arrivals.  It is seen in Fig. 1 that our ray simulation in
the presence of internal waves underestimates the vertical spread of the
finale region.  Simulations with a stronger internal wave field,
$E = 1.0 \: E_{GM}$, give a better fit to the vertical spread of energy (but a
poorer fit to other wavefield features, as described below).  Note, however,
that diffractive effects, missing in our ray simulations, will contribute to
enhanced broadening in depth of the arrival pattern.   A striking feature of
Figs. 1, 2 and 4 is the contrast between the seemingly structureless
distribution of steep (large $|p|$) rays in $(p,z)$ seen in Fig. 4 when
internal waves are present and the relatively structured distribution of the
same rays in the time-depth (Fig. 1) and time-angle (Fig. 2) plots. 

The cause of the complexity seen in Fig. 4 is ray chaos \cite{SVZ,rdoa}; most
ray trajectories diverge from initially infinitesimally perturbed rays at an
exponential rate.  48,000 rays with launch angles between $\pm12^o$ were
traced to produce Figs.  1, 2 and 4 in the presence of internal waves.  This
fan of rays is far too sparse to resolve what should be an unbroken smooth
curve -- a Lagrangian manifold -- which does not intersect itself in phase
space (Fig. 4).  Under chaotic conditions the separation between neighboring
rays grows exponentially, on average, in range.  The complexity of the
Lagrangian manifold grows at the same exponential rate.  The Lyapunov exponent
is the recriprocal of the e-folding distance (see, e.g., \cite{rdoa}).  Finite
range numerical estimates of Lyapunov exponents (hereafter referred to as
stability exponents) are shown as a function of launch angle in Fig. 5.  It
is seen that in this environment the flat rays ($|\varphi_0| \leqa 5^o$) have
stability exponents of approximately 1/(100 km), while the steeper rays
($6^o \leqa |\varphi_0| \leqa 11^o$) have stability exponents of approximately
1/(300 km).  It follows, for example, that the complexity of the flat ray
portion of the Lagrangian manifold grows approximately like
$\exp(r/100 \; {\rm km})$. 

Not surprisingly, stability exponents show some sensitivity to parameters
that describe the internal wave field.  Our simulations show that the stability
exponents of rays with axial angles of approximately $5^o$ or less
approximately double when $j_{max}$ increases from 30 to 100, while steeper
rays show very little sensitivity to $j_{max}$.  In contrast, stability
exponents of rays with axial angles from approximately $8^o$ to $15^o$ 
approximately double when $k_{max}$ increases from $2\pi/1000$ m to
$2\pi/250$ m, while the flatter rays show little sensitivity to $k_{max}$.
With these comments in mind, one should not attach too much significance to
the details of Fig. 5.  We note, however, that rays with axial angles of
approximately $5^o$ or less consistently have higher stability exponents than
rays in the $5^o$ to $10^o$ band.

Note that what appears to be temporally resolved arrivals in the measured
wavefield correspond in our simulations to contributions from an exponentially
large number of ray paths.  (Only a small fraction of the total number are
included in our simulations, however, because of the relative
sparseness of our initial set of rays.)  The observation that the travel
times of chaotic ray paths may cluster and be relatively stable was first
made in Ref. \cite{PGJ}.

Simmen et al. \cite{SFW} have previously produced plots similar to our Figs. 1
and 4 for ray motion in a deep ocean model consisting of a background sound
speed structure very similar to ours on which internal-wave-induced sound speed
perturbations were superimposed.  Although our results are similar in many
respects, it is noteworthy that the right panel of our Fig. 4 shows more
chaotic behavior than is present in the corresponding plot in Ref. \cite{SFW}.
This difference persists if our 3252.38 km range simulations are replaced
by simulations at the same range (1000 km) that was used in Ref. \cite{SFW}.
Our rays, especially the steeper rays, are more chaotic than those in Ref.
\cite{SFW}.  We believe that this difference is primarily due to the increased
complexity of our internal wave field over that of Simmen et al. \cite{SFW} who
included contributions from ten internal wave modes in their simulated
internal wave fields.

Figures 6 and 7 show plots of ray depth at the AET range vs. launch angle.
In Fig. 6 the same rays (excluding those with negative launch angles) that
were used to produce Figs. 1, 2 and 4 are plotted.  In Fig. 7 two small
angular bands of the upper panel of Fig. 6 are blown up; the ray density is
four times greater than that used in Fig. 6.  Fig. 6 strongly
suggests that, even in the absence of internal waves, the near-axial rays
are chaotic.  Not surprisingly, in the presence of internal waves, steeper
rays are also predominantly chaotic.  Note, however, that Figs. 5, 6 and 7
show that the near-axial rays are evidently more chaotic than the steeper
rays.  An explanation for this difference in behavior will be given in
section III. 

In this section we have briefly described and compared the gross features of
measured and simulated AET wavefields.  Ray trajectories in our simulated
wavefields in the presence of internal waves are predominantly chaotic.  In
spite of this, Figs. 1 and 2 show that many features of our simulated
wavefields are stable and in good qualitative agreement with the observations.
Predicted and measured spreads of
acoustic energy in time, depth and angle are generally in good agreement, both
in the early and late portions of the arrival pattern.  We shall not discuss
further the spreads of energy in depth and angle, except to note that Figs.
1 and 2 show that these spreads can be accounted for using ray methods
in the presence of realistic (including internal-wave-induced sound speed
perturbations) ocean structure.  In the two sections that follow we shall
consider in more detail the time spreads of the early ray arrivals, and
intensity statistics in both the early and late portions of the arrival
pattern. 

\section{Chaos, micromultipaths and timefront stability}

In this section we consider in some detail eigenrays, timefront stability and
time spreads.  We focus our attention on the early branches of the timefront,
where measured time spreads were, surprisingly, only approximately 2 ms, in
sharp contrast to the theoretical prediction of approximately 1 sec \cite{AET2}.
We focus our attention on the early arrival time spreads, primarily because
quantitative estimates of the unresolved late arrivals are not available.

Eigenrays at a fixed depth $z$ correspond to the intersections of a horizontal
line at depth $z$ with the curve $z(\varphi_o)$; these intersections are the
roots of the equation $z - z(\varphi_o) = 0$.  Although the discrete samples
of $z(\varphi_o)$ plotted in Figs. 6 and 7 are too sparse to reveal what should
be a smooth curve, it is evident the number of eigenrays at almost all depths
is very large; because rays are predominantly chaotic, this number grows
exponentially in range.  In principle, eigenrays can be found using a
combination of interpolation and iteration, starting with a set of discrete
samples of $z(\varphi_o)$.  In practice, this procedure reliably finds only
those eigenrays in regions where $z(\varphi_o)$ has relatively little
structure.  These rays have the highest intensity and are the least chaotic.
It is seen in Fig. 1 that the early
(steeper) ray arrivals come in clusters with small travel time spreads.
The set of eigenrays within such a cluster is often referred to as a set of
micromultipaths.  An incomplete set of micromultipaths, found using the
procedure just described, is shown in Fig. 8.  An interesting and important
feature of micromultipaths in our simulations is that all of the rays that
make up a set of micromultipaths have the same identifier $\pm M$.  Here
$\pm$ is the sign of the launch angle and $M$ is the number of ray turning
points (where $p$ changes sign) following a ray.  In Fig. 9 rays with two
fixed values of ray identifier (+137 and +151) are plotted in the time-depth
plane.  In both cases the points plotted are a subset of those plotted in
Fig. 1.  It is seen that the relatively steep +137 rays have a small time
spread and form one of the branches of the timefront seen in Fig. 1
(similar behavior was noted in Ref. \cite{SFW}), while the relatively flat
+151 rays have a much larger time spread and fall within the diffuse finale
region of the arrival pattern seen in Fig. 1.

An important (and surprising) feature of sets of micromultipaths is that they
are highly nonlocal in the sense that interspersed (i.e., having intermediate
launch angles) among a group of micromultipaths are rays with $M$ values that
differ by several units.  This behavior is illustrated in Fig. 10.  In view
of the observation that, in the presence of internal waves in the AET
environment, ray motion is strongly chaotic and the function $M(\varphi_o)$ has
local oscillations of several units, it is remarkable that the early portion
of the timefront (see Fig. 1) is not destroyed.  The nonlocality of a set of
micromultipaths in the range-depth plane is seen in Fig. 8.  Note that there
are significant differences in the upper and lower turning depths of the
plotted micromultipaths and that these rays are spread in range by a
significant fraction of a ray cycle distance.

The result of performing a ray-based synthesis of what appears in the
measurements to be an isolated arrival is shown in Fig. 11.  This involves
finding a complete set of micromultipaths, including their travel times and
Maslov indices, and coherently summing their contributions.  Unfortunately,
owing to the
predominantly chaotic motion of ray trajectories in the AET environment in the
presence of internal waves, it is extremely difficult to find a complete set
of micromultipaths.  Indeed, the constituent ray arrivals used in the Fig. 11
synthesis do not constitute a complete set of micromultipaths.  This is less
important than might be expected because standard eigenray finding techniques
easily locate the strongest micromultipaths; the missing micromultipaths in
the Fig. 11 synthesis are highly chaotic rays that have very small amplitudes.
An interesting result of this synthesis is that the micromultipath-induced
time spread in the synthesized pulse is only about 1 ms, which is in very good
agreement with the AET measurements \cite{AET2,CDT}.  Note that the total
time spread among the micromultipaths found -- about 11 ms -- is much greater
than the spread of the synthesized pulse.  This difference arises because the
time spread of the dominant micromultipaths is much smaller than the total
micromultipath time spread.

Simulations (not shown) using $E = 1.0 \: E_{GM}$ result in synthesized steep
ray pulses that are spread in time much more than is shown in Fig. 11.  With
the stronger internal wave field, the total micromultipath time spread
increases by approximately a factor of two -- to about 25 ms.  More importantly,
however, the dominant micromultipaths have time spreads comparable to the
total micromultipath time spread, leading to synthesized pulses that are
spread by 5 to 10 ms, rather than 1 to 2 ms.  Thus, measured early arrival
time spreads are in better agreement with $E = 0.5 \: E_{GM}$ simulations
than with $E = 1.0 \: E_{GM}$ simulations.  The question of which value of
the internal wave strength $E$ in our simulations gives the best fit to the
observations will be revisited in the next section.

An interesting and unexpected feature of our simulations is that the Maslov
index was consistently lower than the number of turns $M$ made by the same
ray.  The difference was typically three units in the $E = 0.5 \: E_{GM}$
simulations and four units in the $E = 1.0 \: E_{GM}$ simulations.  We have
no explanation for this behavior.

So far in this section we have seen that associated with the chaotic motion
of ray trajectories is extensive micromultipathing, and that the
micromultipathing process is highly nonlocal.  The many micromultipaths that
add at the receiver to produce what appears to be a single arrival may sample
the ocean very differently.  Surprisingly, on the early arrival branches the
highly nonlocal micromultipathing process causes only very small time spreads
and does not lead to a mixing of ray identifiers.  To complete this picture
it is necessary to explain why the time spreads of the early (steep) ray
arrivals are so small.  We shall now provide such an explanation.  Recall
first, however, that the measured time spread of the early arrivals is only 
approximately 2 ms \cite{AET2}, in marked contrast to the theoretical
prediction \cite{FDMWZ,AET2} of approximately 1 sec.  Thus our quantitative
explanation for the small early arrival time spreads serves both to give
insight into the underlying physics, and provides an explanation for a 
critically importnat feature of the measured AET wavefields.

Computing time spreads is conceptually straightforward using ray methods.
In a known environment one finds a large ensemble of rays with the same
identifier that solve the same two point boundary value problem; the spread
in travel times over many ensembles of such rays, each ensemble computed
using in a different realization of the environment, is the desired quantity.
The constraint that all of the computed rays have the same fixed endpoints
complicates this calculation.  In the following we exploit an approximate form
of the eigenray constraint that simplifies the calculation; the approximate
form of the eigenray constraint is first order accurate in a sense to be
described below.

Before describing the manner in which the eigenray constraint is imposed, it
is useful to describe the physical setting in which the constraint will be
applied.  Figure 12 shows $|H|$ vs $r$ following a moderately steep ray (axial
angle approximately $11^o$) in a canonical environment \cite{Mcan} on which
the previously
described (based on the measured AET $N(z)$ profile) internal-wave-induced
sound speed perturbation field was superimposed.  It is seen that $H(r)$
consists of a sequence of essentially constant $H$ segments, separated by
near-step-like jumps.  The jumps occur at the ray's upper turning points.  A
simple model that captures this behavior -- the so-called apex approximation
\cite{FDMWZ} -- assumes that the transition regions can be approximated as
step functions.  (The validity of the apex approximation is linked to the
anisotropy and inhomogeneity of oceanic internal waves; details of the
background sound speed profile are not important.  For this reason we have
chosen to illustrate
this effect using a background canonical profile rather than the AET
environment.)  Figure 13 shows a schematic diagram of two rays.  One ray
has undergone two internal-wave-induced apex scattering events; the other
ray is an unscattered ray with the same launch angle as the scattered ray.

Our strategy to building the eigenray constraint into an estimate of travel
time spreads can now be stated.  We consider a scattered and an unperturbed
(that is, in the absence of internal-wave-induced scattering events) ray
with the same ray identifier; the rays shown schematically in Fig. 13 have
identifier $+3$.  In addition to having the same ray identifier, the scattered
and unperturbed rays are assumed to start from the same point and end at the
same depth, but they will generally end at different ranges.  We assume that
the scattered ray has travel time $T_r$ and is one of many scattered eigenrays
with the same ray identifier at range $r$.  We estimate the travel time $T(r)$
of an unperturbed eigenray at range $r$ whose ray identifier is the same as
that of the scattered eigenray.  It will be shown that
$\delta T(r) = T_r - T(r)$ vanishes to first order if the apex approximation
is exploited.  Because the same result applies to all of the scattered
eigenrays, it follows that the time spread vanishes to first order, independent
of $r$, if the apex approximation is exploited.  Note that this result is not
expected to apply, even approximately, to the
late near-axial rays where the apex approximation is known to fail.

The travel time $T(r)$ of the eigenray in the unperturbed ocean can, of
course, be computed numerically but this provides little help in deriving an
analytical estimate of $\delta T$.  Instead, consider a ray in the unperturbed
ocean with the same launch angle as one of the scattered eigenrays, as shown
in Fig. 13.  In general, the range of this ray at the receiver depth, after
making the appropriate number of turns, will be $r_o \neq r$.  But $T(r)$
can be estimated from $T(r_o)$.  This follows from the observation that ray
travel time (assuming a point source initial condition, for instance) is a
continuously differentible function of $z$ and $r$ with $\nabla T = \bf{p}$
where $\bf{p}$ is the ray slowness vector.  Because of this property $T(z,r)$
can be expanded in a Taylor series.  If we consider rays with the same ray
identifier and fix the final ray depth to coincide with the receiver depth,
then
\begin{equation}
T(r) \approx T(r_o) + p_r(r_o)(r - r_o)
\label{T(r)}
\end{equation}
where $p_r = -H$ is the $r$-component of the ray slowness vector.  (More
generally, $T(z,r)$ consists of multiple smooth sheets that
are joined at cusped ridges; in spite of this complication, the property of
continuous differentiability is maintained, even in a highly structured
ocean.  For our purposes, we need only consider one sheet of this multi-valued
function.  Note also that our application of (\ref{T(r)}) in the background
ocean makes this expression particularly simple to use.)  It follows from
(\ref{T(r)}) that
\begin{equation}
\delta T(r) = T_r - T(r) \approx T_r - T(r_o) - p_r(r_o)(r - r_o)
\label{deltaT}
\end{equation}
is, correct to first order, the travel time difference at range $r$ between
eigenrays with and without internal waves.

To evaluate $\delta T(r)$ we shall assume that
the backgound environment is range-independent.  Also, consistent with the
use of the apex approximation, we may treat the perturbed environment
as piecewise range-independent.  Within each range-independent segment it is
advantageous to make use of the action-angle variable formalism (see. e.g.,
\cite{rdoa} or \cite{AV01}).  In terms of the action-angle variables
$(I,\theta)$, $\tilde{H} = \tilde{H}(I)$ and the ray equations are
\begin{equation}
\frac{d\theta}{dr} = \frac{\partial \tilde{H}}{\partial I} = \omega(I),
\label{aa1}
\end{equation}
\begin{equation}
\frac{dI}{dr} = - \frac{\partial \tilde{H}}{\partial \theta} = 0,
\label{aa2}
\end{equation}
and
\begin{equation}
\frac{dT}{dr} = I\frac{d\theta}{dr} - \tilde{H}(I) = I\omega(I) - \tilde{H}(I),
\label{aa3}
\end{equation}
where the physical interpretation of $\tilde{H}$ as $-p_r$ is maintained.  The
angle variable can be defined to be zero at the upper turning depth of a ray
and increases by $2\pi$ over a ray cycle (double loop).  It follows that the
frequency $\omega(I) = 2\pi/R(I)$ where $R(I)$ is a ray cycle distance.
Integrating the ray equations over a complete ray cycle then gives
$R(I) = 2\pi/\omega(I)$ and $T(I) = 2\pi(I - \tilde{H}(I)/\omega(I))$ with $I$
constant following each ray.  In the apex approximation $I$ jumps
discontinuously at each ray's upper turning depth; for the perturbed ray
$R(I + \Delta I) \approx 2\pi/\omega(I) - 2\pi\Delta I \omega'(I)/(\omega(I))^2$
and $T(I + \Delta I) \approx 2\pi(I - \tilde{H}(I)/\omega(I)) +
2\pi\Delta I \tilde{H}(I)\omega'(I)/(\omega(I))^2$ where
$\omega'(I) = d\omega(I)/dI$.  Note that, like these expressions, Eqs.
\ref{T(r)} and \ref{deltaT} are first order accurate in $\Delta I$.

Consider again Eq. \ref{deltaT} and Fig. 13 with $I$ equal to $I_0$, $I_1$
and $I_2$ in the left center and right ray segments, respectively.  The left
segment gives no contribution to $\delta T$ as the ray has not yet been
perturbed.  In the center ray segment
\begin{equation}
T_r - T(r_o) \approx
2\pi(I_1 - I_0)\frac{\tilde{H}(I_0)\omega'(I_0)}{(\omega(I_0))^2}
\label{delt}
\end{equation}
and
\begin{equation}
-p_r(r_o)(r - r_o) \approx \tilde{H}(I_0)
     \left[ -2\pi(I_1 - I_0)\frac{\omega'(I_0)}{(\omega(I_0))^2} \right].
\label{pdelr}
\end{equation}
These terms are seen to cancel.  Note that if additional complete cycle ray
segments are added to the center section of the ray, Eqs. \ref{delt} and
\ref{pdelr} are unchanged except that $I_1 - I_0 = \Delta I_1$ is replaced by
$\sum_{i=1}^n \Delta I_i = I_n - I_0$; again the two terms cancel.  In the
final (incomplete cycle) ray segment the difference between the terms
$T_r - T(r_o)$ and $p_r(r_o)(r - r_o)$ can be shown to be $O((\Delta I)^2)$;
the terms do not exactly cancel because the $\theta$-values of the perturbed
and unperturbed rays are generally not identical at the receiver depth. Thus,
to first order in $\Delta I$, $\delta T = 0$, independent of range, if the
apex approximation is valid.  This simple calculation provides an explanation
of why, in spite of extensive ray chaos, the time spreads of the early AET
ray arrivals are quite small.

Several comments concerning the preceeding calculation are noteworthy.  First,
we note that although it was assumed that the background sound speed structure
is range-independent, the preceeding argument also holds in the presence of
slow background range-dependence, i.e., with structure whose horizontal scales
are large relative to a typical ray double loop length.  Adiabatic invariance
in such environments guarantees that while $H$ is not constant following rays
between apex scattering events, $I$ is nearly constant.  Second, after $n$
random kicks $I_1 - I_0$ in (\ref{delt}) and (\ref{pdelr}) is replaced by
$I_n - I_0 \approx \sqrt{n}(\Delta I)_{rms}$, and the magnitude of (\ref{delt})
under AET conditions is $O(1 \; \rm{sec})$.  This is an example of a travel
time spread estimate that fails to enforce the eigenray constraint.   This
calculation shows that the difference between travel time estimates that do
and do not enforce the eigenray constraint can be quite significant.  (The
constrained estimate is refined below.) Third, it should be emphasized that
Eqs. (\ref{delt}) and (\ref{pdelr}) apply (approximately) only to the steep
rays because the apex approximation applies only to the steep rays.  We have
not addressed the time spreads of near-axial rays.  Note, however, that Fig. 9
shows that time spreads are larger for flatter rays.  Fourth, the above
calculation shows that, to lowest order in $\Delta I$, there is no
internal-wave-scattering-induced travel time bias if the apex approximation
is strictly applied.  And fifth, the arguments leading to Eqs. (\ref{delt})
and (\ref{pdelr}) apply whether the scattered rays are chaotic or not.

A relaxed form of the apex approximation in which the action jump
transition region has width $\Delta \theta$ gives a nonzero travel time spread
estimate.  In the transition region, taken for convenience to be
$0 \leq \theta \leq \Delta \theta$, one may choose a Hamiltonian of the form
$\tilde{H} = \tilde{H}(I - s \theta)$ where $s = \Delta I/\Delta \theta$, and
$\tilde{H} = \tilde{H}(I)$ elsewhere.  It follows that in the transition
region $I(\theta) = I_0 + s \theta$ and $I = \rm{constant}$ elsewhere.  A
simple generalization of the above calculation then gives for a complete ray
cycle
\begin{eqnarray}
 T_r - T(r_o)  \approx & &  \nonumber \\
& & \!\!\!\!\!\!\!\!\!\!\!\!\!\!\!\!\!\!\!\!\!\!\!\!\!\!\!\!\!\! 
(I_1 - I_0) \left [
 (2\pi - \Delta \theta)
\frac{\tilde{H}(I_0)\omega'(I_0)}
{(\omega(I_0))^2} + \frac{\Delta \theta}{2} \right ] 
\label{delt2}
\end{eqnarray}
and
\begin{eqnarray}
 -p_r(r_o)(r - r_o) \approx & &  \nonumber \\
& & \!\!\!\!\!\!\!\!\!\!\!\!\!\!\!\!\!\!\!\!\!\!\!\!\!\!\!\!\!\! 
\tilde{H}(I_0)
\left[ -(2\pi - \Delta \theta)(I_1 - I_0)\frac{\omega'(I_0)}{\omega(I_0)^2}
\right],
\label{pdelr2}
\end{eqnarray}
so the sum (recall [\ref{deltaT}]) is
\begin{equation}
\delta T \approx \frac{\Delta \theta}{2}(I_1 - I_0). \\
\label{deltaT2}
\end{equation}
It should be emphasized that this expression is first order accurate in
$\Delta I = I_1 - I_0$, but that no assumption about the smallness of
$\Delta \theta$ has been made.  As noted above, incomplete cycle end segment
pieces give $O((\Delta I)^2)$ corrections to $\delta T$.  If $\Delta \theta$
has approximately the same value at each upper turn, then one has after $n$
upper turns, correct to $O(\Delta I)$,
\begin{eqnarray}
\delta T
 \approx \frac{\Delta \theta}{2} \sum_{i=1}^n(I_i - I_{i-1})
 & = & \frac{\Delta \theta}{2} (I_n - I_0) \nonumber \\
&  \approx & \frac{\Delta \theta}{2} \sqrt{n} (\Delta I)_{rms}. 
\label{deltaTsum}
\end{eqnarray} 
Consistent with the numerical simulations shown in Fig. 12,
$\Delta \theta \approx 0.8$ radians and $\Delta I \approx 4$ ms.  (We, and
independently F. Henyey [personal communication], have confirmed that these
estimates also apply under AET-like conditons.)  With these
numbers and $\sqrt{n} = 8$, appropriate for AET, (\ref{deltaTsum}) gives a
time spread estimate of approximately 13 ms, in approximate agreement with
the numerical simulations shown in Fig. 11.  Note that (\ref{deltaTsum}) does
not preclude a travel time bias.  A cautionary remark concerning the use of
(\ref{deltaTsum}) is that Virovlyansky \cite{XX} has pointed out that, owing
to secular growth, the $O((\Delta I)^2)$ contribution to $\delta T$ may
dominate the $O(\Delta I)$ contribution at long range.

Figure 9 shows that in the AET environment simulated near-axial ray time
spreads are greater than simulated steep ray time spreads.  We do not fully
understand this behavior.  A possible explanation for this behavior is that
time spreads increase as rays
become increasingly flat owing to the breakdown of the apex approximation.
But one would expect that this trend should be offset, in part or whole, by
the relative smallness of internal-wave-induced sound speed perturbations
near the sound channel axis.  In addition, we have seen some evidence that an
additional factor may be important in the AET environment.  Namely, we have
observed a positive correlation between travel time spreads and stability
exponents; stability exponents in the AET environment are shown in Fig. 5.
We have chosen not to dwell on flat ray time spreads in this paper because
there are no AET measurements of these spreads to which simulations can be
compared.  It is clear, however, that the issues just raised need to be
better understood.

\section{Wavefield intensity statistics}

In this section we consider the statistical distribution of the intensities
of both the early and late AET arrivals.  Recall that experimentally the early
arrival intensities have been shown \cite{AET2,CDT} to approximately fit a
lognormal probability density function (PDF) and the late arrival intensities
have been shown \cite{CDT} to fit an exponential distribution.
The late arrival exponential distribution is not surprising as this
distribution is characteristic of saturated statistics.  The early
arrival near-lognormal distribution is surprising, however, inasmuch as
theory \cite{FDMWZ,AET2} predicts saturated statistics, i.e., an exponential
intensity PDF.  It has been argued \cite{C99} that the theory can be modified
in such a way as to move the early arrival prediction from saturated
to unsaturated statistics.  The latter regime is characterized by a lognormal
intensity PDF.  This fix is conceptually problematic inasmuch as, in this
theory, the unsaturated regime is characterized by the absence of
micromultipaths, which seriously conflicts with the numerical simulations
presented in the previous section where the number of micromultipaths is very
large.  In this section we provide self-consistent explanations for both the
late arrival exponential distribution and the early arrival lognormal
distribution.  The challenge is to reconcile the early arrival near-lognormal
intensity PDF with the presence of a large number of micromultipaths.  Some
of the arguments presented are heuristic, and some build on the results of
numerical simulations.  A complete theoretical understanding of intensity
statistics has proven difficult.

Our approach to describing wavefield intensity statistics builds on the
semiclassical construction described by Eq. (\ref{geom}).  At a fixed location
it is seen that the wavefield amplitude distribution is determined by the
distribution of ray amplitudes and their relative phases.  Note that both
travel times and Maslov indices influence the phases of ray arrivals.
Complexities associated with transient wavefields and caustic corrections
will be discussed below. 

An important observation is that in the AET environment, including
internal-wave-induced sound speed perturbations, simulated geometric amplitudes
of both steep and flat ray arrivals approximately fit lognormal PDFs.
This is shown in Fig. 14.  Previously it has been shown \cite{WTo} that ray
intensities in a very different chaotic system also fit a lognormal PDF; in
that system single scale isotropic fluctuations are superimposed on a
homogeneous background.
(Note that all powers of a lognormally distributed variable are also
lognormally distributed, so ray amplitudes have this property if and only if
ray intensities have this property.)  The apparent generality of the
near-lognormal ray intensity PDF suggests that it applies generally to ray
systems that are far from integrable; the arguments presented in Ref.
\cite{WTo} suggest that this should be the case.  The intensity distributions
of the early and late AET arrivals, corresponding to steep and flat rays,
respectively, will be discussed separately.

We consider first the early arrivals.  We saw in the previous section that
the micromultipaths that make up one of these arrivals have the same ray
identifier and have a very small spread in travel time.  The dominant
micromultipaths were seen (see Fig. 11) to have a time spread of approximately
1 ms which is a small fraction of the approximately 13 ms period of the 75
Hz carrier wave.  Also, our simulations show that the Maslov indices of the
dominant micromultipaths differ by no more than one unit.  These conditions
dictate that interference among the dominant micromultipaths is predominantly
constructive.  Because travel time differences are so small, the pulse shape
should have negligible influence on the distribution of peak intensities.
In a model of the early AET arrivals consisting of a superposition of
interfering micromultipaths, the micromultipath properties that play a critical
role in controlling peak wavefield intensities are thus: 1) their amplitudes
have a near-lognormal distribution; 2) the dominant micromultipaths have
negligible travel time differences; and 3) the dominant micromultipaths have
Maslov indices that differ by no more than one unit.  It should be noted that
very different behavior would have been observed if: the source bandwidth
were significantly more narrow as this would have caused micromultipaths with
different ray identifiers to interfere with one another; or the source center
frequency were significantly higher as phase differences between interfering
micromultipaths would then have been significant. 

An additional subtlety must be introduced now: the PDF of the intensities of
the constituent micromultipaths that make up a single arrival is not identical
to the PDF described in Ref.~\cite{WTo} and shown in the upper and middle
panels of Fig. 14.  The latter PDF describes the distribution of intensities
of randomly (with uniform probability) selected rays leaving the source within
some small angular band and whose range is fixed.  This PDF is biased in the
sense that it overcounts the micromultipaths with large intensities and
undercounts those with small intensities.  Unbiased micromultipath intensity
PDFs can be constructed from the biased PDFs shown.  To do so, consider a
manifold (a smooth curve in phase space corresponding to a fan of initial
rays) which begins near $(z_0,p_0)$ and arrives in the neighborhood of
$(z_r,p_r)$ at range $r$.  Sampling in fixed steps of the differential
$\delta p_0$ (as was done to produce the upper and middle panels of Fig. 14;
this is equivalent to uniform random sampling) leads to a highly nonuniform
density of points on the final manifold since
$(z_0,p_0+\delta p_0)$ propagates to $(z_r+q_{21} \delta p_0, p_r+q_{11}
\delta p_0)$.  The greater $q_{21}$, the lower the density of points
locally at final range.  A uniform sampling in final position is acheived
instead by considering the initial conditions that would lead to
$(z_r+\delta z_r, p_r+q_{11}/q_{21} \delta z_r)$.  Its density of points
on the initial manifold can be deduced from its initial condition, $(z_0,
p_0+\delta z_r/q_{21})$.  It is necessary to sample $q_{21}$ times more
densely on the initial manifold in order to acheive uniform sampling in
$\delta z_r$ at range $r$.  To account for this effect, we need
to know the PDF for $q_{21}$ with uniform initial sampling.  Roughly
speaking, the PDF of the absolute values of the individual matrix
elements of $q$ have the same form as for $|Tr(Q)|$, apart from a shift of
the centroid that is lower order in range than the leading term.  From
the results of Ref.~\cite{WTo}, the (biased in the sense described above)
probability that $q_{21}$ falls in the interval between $x$ and $x + dx$ is
\begin{eqnarray}
\label{rhotrm}
\rho_{|q_{21}|}(x)=\sqrt{\frac{1}{2\pi r(\bar\nu-\nu_L)}} \frac{1}{ x}
\; \cdot \qquad & & \nonumber \\
& & \!\!\!\!\!\!\!\!\!\!\!\!\!\!\!\!\!\!\!\!\!\!\!\!\!\!\!\!\!\! 
\!\!\!\!\!\!\!\!\!\!\!\!\!\!\!\!\!\!\!\!\!\!\!\!\!\!\!\!\!\! 
\exp{\left [ \frac{- \left  ( \ln{(x)}-\nu_L r \right)^2}
{2r(\bar\nu-\nu_L)} \right ]}  \; , \; x \geq 0 \; .
\end{eqnarray}
Here $\bar\nu$ is a finite-range estimate of the Lyapunov exponent based
on an average (over an ensemble of rays) value of $q_{21}$, and $\nu_L$
is the true Lyapunov exponent.
The new (unbiased in the sense described above) PDF, $\rho^\prime$, for
uniform sampling in $\delta z_r$ is related to the previous one by
\begin{eqnarray}
\label{rhotrm2}
\rho^\prime_{|q_{21}|}(x) &=& \frac{x}{\langle x\rangle}
\rho_{|q_{21}|}(x) \nonumber \\
&=& \sqrt{\frac{1}{2\pi r(\bar\nu-\nu_L)}} \frac{1}{ x} 
\; \cdot \nonumber \\
& & 
\exp{\left [ \frac{- \left  ( \ln{(x)}-\bar\nu r \right)^2}
{2r(\bar\nu-\nu_L)} \right ]}  \; , \; x \geq 0
\end{eqnarray}
where the factor $x$ accounts for the extra counting weight of $q_{21}$,
and $\langle x\rangle$ just preserves normalization and is calculated
using Eq.~(\ref{rhotrm}).  This calculation shows that the unbiased
micromultipath ray intensity PDF also has a lognormal distribution; the only
change relative to the biased PDF is an increase in the mean from $\nu_L r$
to $\bar\nu r$.  Because lognormality is maintained, this correction represents
only a trivial change to the problem.  

Approximate lognormality of the constrained (eigenray) PDF of ray intensity
is shown in the lower panel of Fig. 14.  This PDF was constructed using the
same eigenrays that were used to produce Figs. 8 and 11.  The corresponding
unconstrained ray intensity PDF is shown in the middle panel of Fig. 14.
(Two constraints -- fixed receiver depth and fixed ray identifier -- are
built into the lower panel PDF.  A very similar  constrained PDF results if
only the receiver depth constraint is applied, provided ray launch angles are
limited to the `steep' ray band used to construct the middle panel.)  It
should be noted that the constrained (eigenray) PDF shown in the lower panel
of Fig. 14 was constructed from numerically found eigenrays; because weak
eigenrays are difficult to find numerically they are undercounted and the
PDF is biased.  In a practical sense this bias is of little consequence because
the weak eigenrays that are difficult to find contribute negligibly to the
wavefield.

We return now to the problem of simulating the early AET arrivals.  It is
tempting to think that because the constituent micromultipaths have a
near-lognormal distribution, the sum of their contributions should also be
near-lognormally distributed.  Unfortunately, this is generally not the case.
Consider, for example, the special case in which phase, including Maslov
index, differences are negligible.  Then all micromultipaths interfere
constructively and peak wavefield amplitudes can be modelled as the sum of
many lognormally distributed variables.  Because all moments of the lognormal
distribution are finite, the central limit theorem applies.  Under these
conditions, if sufficiently many contributions are summed, the distribution
of the sums -- the wavefield amplitudes -- would be a gaussian.

To simulate the statistics of the early AET arrivals (recall Fig. 14 and the
accompanying discussion) we have used several variations of a simple model.
An arrival was modelled as a sum of $n_m$ interfering micromultipaths whose:
1) amplitudes are lognormally distributed; and 2) phases,
$\omega T_i - \mu_i \pi/2 \bmod 2\pi$, have a clearly identifiable peak.
Micromultipath contributions were coherently added.  The peak intensity of
the sum -- whose travel time is not known a priori -- was then recorded.
Using an ensemble of $10^4$ peak intensities, a peak intensity PDF was then
constructed.  Peak intensity PDFs constructed in this fashion were found to
be very close to lognormal; a typical example is shown in Fig. 15.  In this
example $f = \omega/2\pi = 75$ Hz, the $T_i$'s were identical,
$\mu_i \in {j,j+1}$ with equal probability (note that choice of the integer
$j$ is unimportant) and $n_m = 5$.  Other combinations of distributions for
$T_i$ (either a Gaussian or the limiting case of a delta distribution),
$\mu_i$ (taken either from ${j,j+1}$ or ${j-1,j,j+1}$ with equal probability),
and the parameter $n_m$ (between 2 and 100) were tested.  These simulations
showed that provided the phase constraint noted above was satisfied, a near
log-normal peak intensity PDF resulted.  

Two points regarding this simple model are noteworthy.  First, this model does
not constitute a theory of wavefield peak intensity statistics, but it does
serve to demonstrate that our ray-based simulations of the early AET arrivals
are consistent with the measured distribution of peak intensities.  Second,
simulations (not shown), performed with $E = 1.0 \: E_{GM}$ yield sets of
dominant micromultipaths that violate assumption 2); phases are uniformly
distributed and summing micromultipath contributions yields a distribution of
peak intensities that is not close to lognormal.  Thus, our simulations suggest
that a near-lognormal distribution of early arrival peak intensity requires a
relatively weak internal wave field.

A complication not accounted for in the preceeding discussion is the presence
of caustics.  At caustics geometric amplitudes (\ref{gampl}) diverge and
diffractive corrections must be applied.  At short range (on the order of the
first focal distance -- a few tens of km in deep ocean environments) we expect
that intensity fluctuations will be dominated by diffractive effects.  The
entire wavefield should be organized by certain high-order caustics which leads
to a PDF of wavefield intensity with long tails \cite{twink,twink2,twink3}.  In
spite of the importance of diffractive effects at short range -- and probably
also at very long range -- we believe that, in the transitional regime
described above, intensity fluctuations are not dominated by diffractive
effects.  This somewhat counterintuitive behavior can be understood by
noting that in the vicinity of caustics the importance of diffractive
corrections to (\ref{gampl}) decreases as the curvature of the caustic
increases.  Under chaotic conditions the curvature of caustics increases,
on average, with increasing range, so the fraction of the total number of
multipaths that require caustic corrections decreases, on average, with
increasing range.  This is true even as the number of caustics grows
exponentially, on average, in range.  This argument leads to the somewhat
paradoxical conclusion that, prior to saturation at least, we expect that the
importance of caustic corrections decreases with increasing range.

We turn our attention now to the late unresolved AET arrivals, corresponding
to the near-axial rays.  Here, time spreads are sufficiently large that
micromultipaths with different ray identifiers are not temporally resolved,
i.e., are not separated in time by more than $(\Delta f)^{-1}$.  At each
$(z,T)$ in the tail of the arrival pattern the wavefield can be modelled
as a superposition of micromultipath contributions with random phases.  The
quadrature components of the wavefield have the form of sums of terms of the
form
\begin{eqnarray}
x_i & = & a_i \cos(\phi_i) \; , \;  \mbox{and} \nonumber \\
y_i & = & a_i \sin(\phi_i)
\label{yq}
\end{eqnarray}
where $\phi_i$ is a random variable uniformly distributed on $[0,2\pi)$.
The distribution of $a_i$ is close to lognormal, but a correction must be
applied to account for pulse shape as many of the interfering micromultipaths
partially overlap in time.  This correction is unimportant inasmuch as the
central limit theorem guarantees that, provided the distributions of $x_i$
and $y_i$ have finite moments, the distributions of sums of $x_i$ and $y_i$
converge to zero mean gaussians.  Thus, wavefield
intensity is expected to have an exponential distribution, consistent with the
observations.  The comments made earlier about caustics apply here as well.

The question of what causes the transition from the structured early portion
of the AET arrival pattern to the unstructured finale region deserves further
discussion.  Recall that the early resolved arrivals have small time spreads
and peak intensities that are near-lognormally distributed, while the finale
region is characterized by unresolved arrivals and near-exponentially
distributed intensities.  In both regions the time spreads and intensity
statistics are consistent with each other inasmuch as in our simulations
a near-lognormal intensity distribution is obtained only when there is a
preferred phase, while the exponential distribution is generated when phases
are random, i.e., when phases are uniformly distributed on $[0,2\pi)$.

With these comments in mind, it is evident that the most important factor in
causing the transition to the finale region is the increase in
internal-wave-scattering-induced time spreads as rays become less steep; as
time spreads increase, neighboring timefront branches blend together and the
phases of interfering micromultipaths get randomized.  The trend toward
increasing
time spreads as rays become less steep is evident in Fig. 9.  The surprising
result is that the scattering-induced time spreads of the steep arrivals are
so small; we have seen that this can be explained by making use of the apex
approximation.  As noted at the end of the previous section, we do not fully
understand the cause of the trend toward larger time spreads as rays flatten.
In the finale region internal-wave-scattering-induced time spreads exceed the
time difference between
neighboring timefront branches that would have been observed in the absence
of internal waves.  Fig. 1 shows that these time gaps decrease as rays become
increasingly flat.  Indeed, this figure shows that even in the absence of
internal waves there would not have been any resolvable timefront branches in
the last half second or so of the arrival pattern.  Some loss of temporal
resolution in the measurements is of course due to the finite source bandwidth,
but without internal waves (or some other type of ocean fluctuations) wavefield
phases would not be random; instead a stable interference pattern would be
observed.

Finally, we note that an interesting feature of our ray simulations is that
the near-axial rays have much higher stability exponents (typically about
$(100 \: {\rm km})^{-1}$) than the steeper rays (typically about
$(300 \: {\rm km})^{-1}$); see Fig. 5.  Also, note that Figs. 6 and 7 strongly
suggest that the near-axial rays in the AET environment are much more chaotic 
than the steeper rays.  The relative lack of stability of the near-axial rays
in the AET environment is, we believe, caused by the background sound
speed structure.  This topic will be discussed in detail elsewhere.  We have
chosen not to focus on this topic here becuse the measured intensity
statistics in the AET finale region are not very sensitive to the near-axial
ray intensity PDF; as noted above, the argument leading to the expectation
that wavefield intensity in the finale region should have an exponential
distribution holds for a very large class of ray intensity distributions.

Finally, we note that an interesting feature of our ray simulations is that
the near-axial rays have much higher stability exponents (typically about
$(100 \: {\rm km})^{-1}$) than the steeper rays (typically about
$(300 \: {\rm km})^{-1}$); see Fig. 5.  Also, note that Figs. 6 and 7 strongly
suggest that the near-axial rays in the AET environment are much more chaotic
than the steeper rays.  The relative lack of stability of the near-axial rays  
in the AET environment is, we believe, caused by the background sound
speed structure.  This topic is discussed in detail in Ref. \cite{BVB}; there 
it is shown that ray stability is largely controlled by a property of the
background (which is assumed to be range-independent) sound speed profile,
$\alpha = (I/\omega)d\omega/dI$.  Large values of $|\alpha|$ are associated
with ray instability.  $\alpha$ vs axial ray angle in the range-averaged AET
environment, and in five 650 km block range-averaged sections of the AET
environment, are shown in Fig. 16.  (The choice of averaging over 650 km
blocks in range is, of course, arbitrary.  Range averaging should be done
locally, however, because the local sound speed structure may be very different
than that which results after averaging over the entire propagation path.)  The
relatively large near-axial ray values of $|\alpha|$ seen in this figure are
consistent with the strongly chaotic nature of these rays seen in Figs. 5, 6 and
7.  Unfortunately, the measured wavefield intensity statistics in the AET finale
region are not very sensitive to the near-axial ray intensity PDF; as noted
above, the argument leading to the expectation that wavefield intensity in the
finale region should have an exponential distribution holds for a very large
class of ray intensity distributions.  Thus, we are not aware of any way that
the AET measurements can be used to test our finding that the near-axial rays
have larger stability exponents, on average, than the steeper rays.

In summary, we believe that the observed near-lognormal PDF of wavefield
intensity for the early resolved AET arrivals is transitional between
fluctuations dominated by caustics at short range and saturation at long
range, where phase differences will be larger.  Surprisingly, for the steep
ray AET arrivals phase differences among the steep ray AET arrivals are very
small so that saturation has not been reached.  The late-arriving AET energy,
on the other hand, is characterized by interfering micromultipaths with random
phases, leading to an exponential PDF of wavefield intensity.  Here, the
underlying near-lognormal PDF of ray intensity is obscured.  We believe
that diffractive effects do not control the intensity fluctuations of either
the early or late AET arrivals.  Although our
theoretical understanding of many of the issues raised in this section is
clearly incomplete, it is worth emphasizing that our ray-based numerical
simulations of intensity statistics, in which ray trajectories are
predominantly chaotic, are in good qualitative agreement with the AET
measurements.  We do not believe that this agreement is accidental.

\section{Discussion and summary}

In this paper we have seen that in the AET environment, including
internal-wave-induced sound speed perturbations, ray trajectories are
predominantly chaotic.  In spite of extensive ray chaos, many features of
ray-based wavefield predictions were shown to be both stable and in good
agreement with the AET measurements.  Predicted and measured spreads of
acoustic energy in time, depth and angle were shown to be in good agreement.
It was shown that associated with the chaotic motion of ray trajectories is
extensive micromultipathing, and that the micromultipathing process is highly
nonlocal; the many micromultipaths that add at the receiver to produce what
appears to be a single arrival may sample the ocean very differently.  It was
shown, surprisingly, that on the early arrival branches the highly nonlocal
micromultipathing process causes only very small time spreads and does not
lead to a mixing of ray identifiers.  A quantitative explanation for the cause
of the very small time spreads of the early ray arrivals was presented.
Partially heuristic explanations for the near-lognormal and exponential
distributions of wavefield intensities for the early and late arrivals,
respectively, were provided.

The fact that one is able to make apparently robust and accurate predictions
of many wavefield features using ray methods under conditions in which ray
trajectories are predominantly chaotic may surprise some readers.  Chaotic
motion is, after all, intimately linked with unpredictability. This apparent
paradox is reconciled by noting that while individual chaotic ray trajectories
are unpredictable beyond some short predictability horizon, distributions of
chaotic ray trajectories may be quite stable and have robust
properties~\cite{ct}.  
Consider, for example, the motion of the ray whose initial conditions are
$z(0) = -652$ m, $\varphi(0) = 8.5^o$, in the AET environment including a known
internal-wave-induced sound speed perturbation.  At $r = 3252$ km, the depth
$z$, angle $\varphi$, travel time $T$, and even the number of turns $M$ made
by this ray are likely to be, for all practical purposes, unpredicatble.  In
contrast, the distribution in $(z,\varphi,T,M)$ of a large ensemble of rays
with $z(0) = -652$ m, $8^o \leq \varphi(0) \leq 9^o$ is very stable in the
sense that essentially the same distribution is seen for any large ensemble of
randomly chosen (with uniform probability) rays inside this initial angular
band.  It is the latter property that allows us to make meaningful
predictions using an ensemble of chaotic rays in a particular realization of
the internal wave field.  In addition, ray distributions with similar
statistical properties are observed using different realizations of the
internal wave field, suggesting that these statistical properties of rays
are robust. 

Our exploitation of results that relate to ray dynamics, including ray chaos,
leads to a blurring of the distinction that is traditionally made between
deterministic and stochastic wave propagation problems.  The ray dynamics
approach emphasizes the distinction between integrable and nonintegrable ray
systems, corresponding to range-independent and range-dependent environments,
respectively.  In generic range-dependent environments at least some ray
trajectories will exhibit chaotic motion.  The oceanographic origin --
mesoscale variability, internal waves or something else -- of the
range-dependent structure is not critical.  An important conceptual insight
that can come only from exploitation of results relating to ray chaos is that
generically phase space is partitioned into chaotic and nonchaotic regions.
This mixture contributes to a combination of limited determinism and constrained
stochasticity. 

For those who wish to exploit elements of determinism in the propagation
physics for the purpose of performing deterministic tomographic inverses, the
results that we have presented have important implications.  We have seen,
for example, that, in spite of extensive ray chaos, many ray-based wavefield
descriptors are stable and predictable, and should be invertible.  A less
encouraging but important observation is that
although the travel times of the steep early arrivals are stable and can be
inverted, the nonlocality of the micromultipathing process limits one's
ability to invert for range-dependent ocean structure.

For those who wish to understand and predict wavefield statistics, our results
also have important implications.  These comments are based on our analysis of
the AET measurements, but are expected to apply to a large class of long-range
propagation problems.  First, we have seen that stable and unstable ray
trajectories coexist and that this influences wavefield intensity statistics.
Second, we have seen that micromultipathing is extensive and highly nonlocal.
The strongly nonlocal nature of the micromultipathing process is significant
because: a) this process cannot be modelled using a perturbation analysis which
assumes the existance of an isolated background ray path; and b) the effects of
this process will not, in general, be eliminated as a result of local smoothing
by finite frequency effects.  Third, we have seen that the appearance of
stochastic effects is not entirely due to internal waves.  We have seen, for
example, that, even in the absence of internal waves, near-axial rays in the
AET environment
are chaotic.  Recall that we noted in the introduction that part of the
motivation for the present study was to understand the reasons underlying the
finding of Colosi et al. \cite{AET2} that the theory described in Ref.
\cite{FDMWZ} fails to correctly predict the AET wavefield statistics.  This
failure is not surprising in view of the observation that this theory is based
on assumptions that either explicitly violate, or lead to the violation of,
all three of the aforementioned properties of the scattered wavefield.  Note
also in this regard that the combination of extensive micromultipathing,
small time spreads, and lognormally distributed intensities that characterizes
the early AET arrivals is not consistent with any of the three propagation
regimes identified in this theory. 

Although we have argued that a ray-based wavefield description which relies
heavily on results relating to ray chaos can account for all of the important
features of the AET measurements, it should be emphasized that our analysis
has some shortcomings.  Recall that our simulations using $E = 0.5 \: E_{GM}$
resulted in steep arrival time spreads and peak intensity statistics in good
agreement with measurements, but that simulations using $E = 1.0 \: E_{GM}$
did not, although direct measurements favor $E = 1.0 \: E_{GM}$.
We have not provided a rigorous argument to establish the
connection between a near-lognormal ray intensity PDF and a near-lognormal
wavefield intensity PDF.  Indeed, a complete theory of wavefield statistics
which accounts for complexities associated with ray chaos is lacking.  This
is closely linked to the subject of wave chaos \cite{rdoa}, which is currently
not well understood.  Also, we have only briefly discussed the spreads of
energy in depth and angle, and more work needs to be done on quantifying
time spreads.

Finally, we wish to remark that the importance of ray methods is not diminished
by recent advances, both theoretical and computational, in the development of
full wave models, such as those based on parabolic equations.  The latter are
indispensible computational tools in many applications.  In contrast, the
principal virtue of ray methods is that they provide insight into the
underlying wave propagation physics that is difficult, if not impossible, to
obtain by any other means.  The results presented in this paper illustrate
this statement.

\ \\

\noindent
{\bf Acknowledgments}

\noindent
We thank Fred Tappert and Frank Henyey for the benefit of discussions on many
of the topics
included in this paper, and the ATOC group (A. B. Baggeroer, T. G. Birdsall,
C. Clark, B. D.  Cornuelle, D. Costa, B. D. Dushaw, M. A. Dzieciuch, A. M. G.
Forbes, B. M.  Howe, D. Menemenlis, J. A. Mercer, K. Metzger, W. Munk, R. C.
Spindel, P. F.  Worcester and C. Wunsch) for giving us access to the AET
acoustic and environmental measurements.  This work was supported by Code 
321 OA of the U.S. Office of Naval Research.

%\newpage

\begin{figure*}
\includegraphics[height=6.0 in]{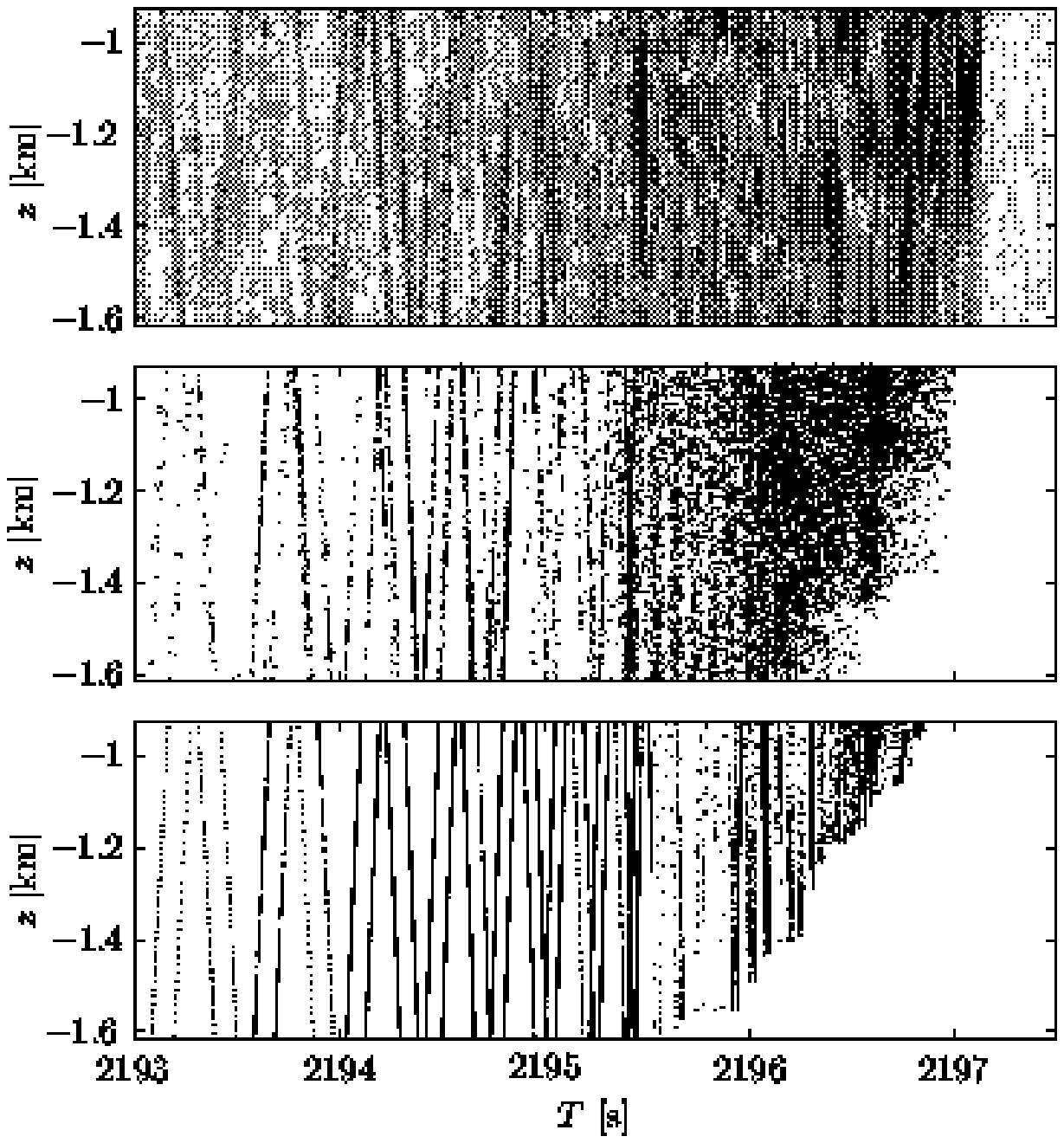} 
\caption{
Measured and simulated AET wavefields in the time-depth plane.  Upper
panel: a typical measurement, shown on a gray scale plot with a dynamic range
of 30 dB, of wavefield intensity.  Middle panel: ray simulation with internal
waves.  Lower panel: ray simulation without internal waves.  Wavefield intensity
in the simulations is approximately proportional to the local density of dots,
each corresponding to a ray.
} 
\end{figure*}

%%%%%%%%%%%%%%%%%%%%%%%%%%%%%%%%%%%%%%%%%%%%%%%%%%%% 
%\newpage
%%%%%%%%%%%%%%%%%%%%%%%%%%%%%%%%%%%%%%%%%%%%%%%%%%%% 
\begin{figure*}
\includegraphics[height=6.0 in]{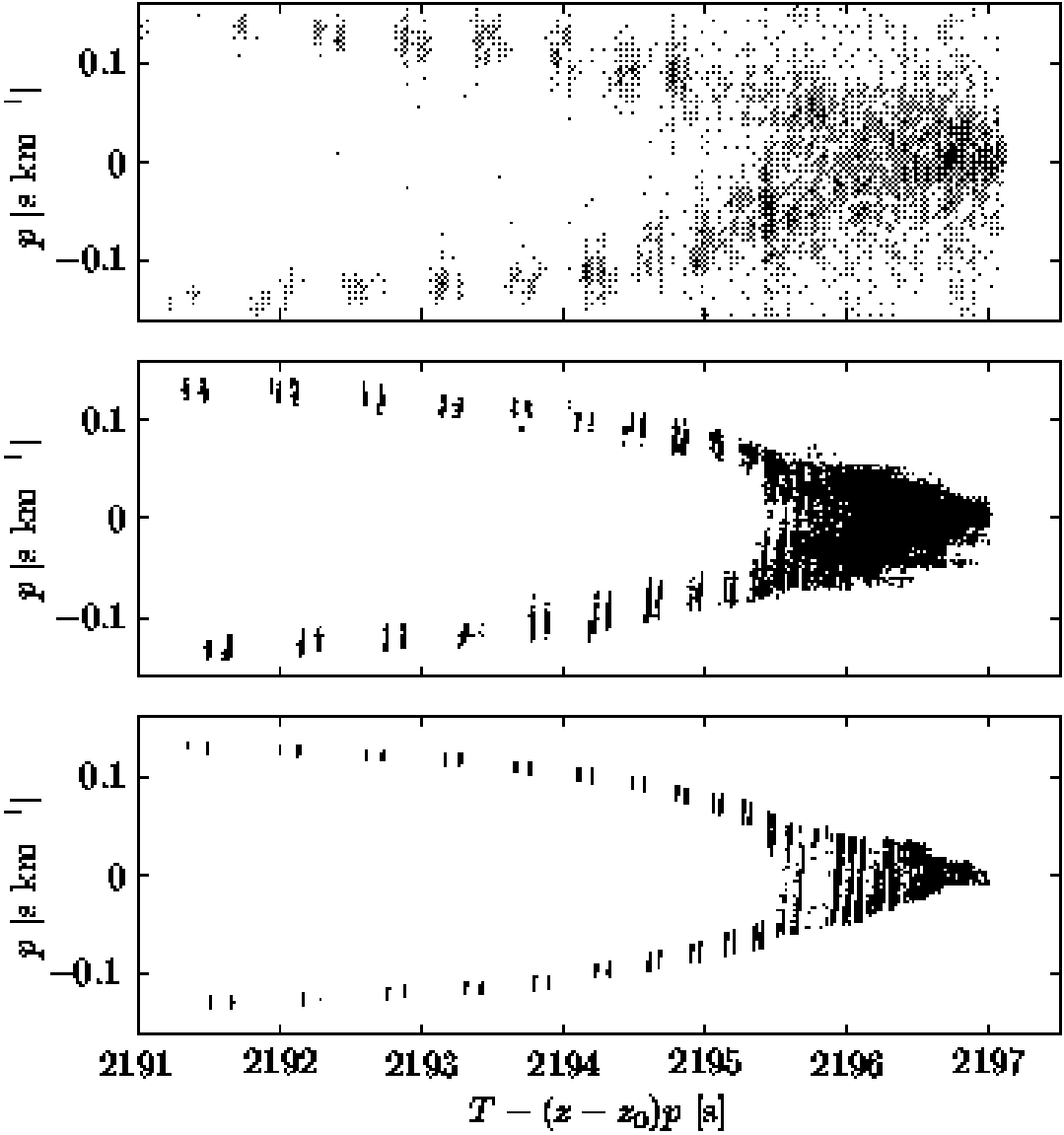}
\caption{
Same as Fig. 1 except that plane wave beamformed wavefields are shown.
The beamformed ray simulations assume a dense receiving array whose upper and
lower bounds coincide with the bounds of the AET receiving array.  The reference
depth was taken to be the depth of the uppermost hydrophone, $z_o = -0.9$ km.
}
\end{figure*}
%%%%%%%%%%%%%%%%%%%%%%%%%%%%%%%%%%%%%%%%%%%%%%%%%%%% 
%%%%%%%%%%%%%%%%%%%%%%%%%%%%%%%%%%%%%%%%%%%%%%%%%%%% 
\newpage

%%%%%%%%%%%%%%%%%%%%%%%%%%%%%%%%%%%%%%%%%%%%%%%% 
%%%%%%%%%%%%%%%%%%%%%%%%%%%%%%%%%%%%%%%%%%%%%%%% 
\begin{figure*}
\includegraphics[height=6.0 in]{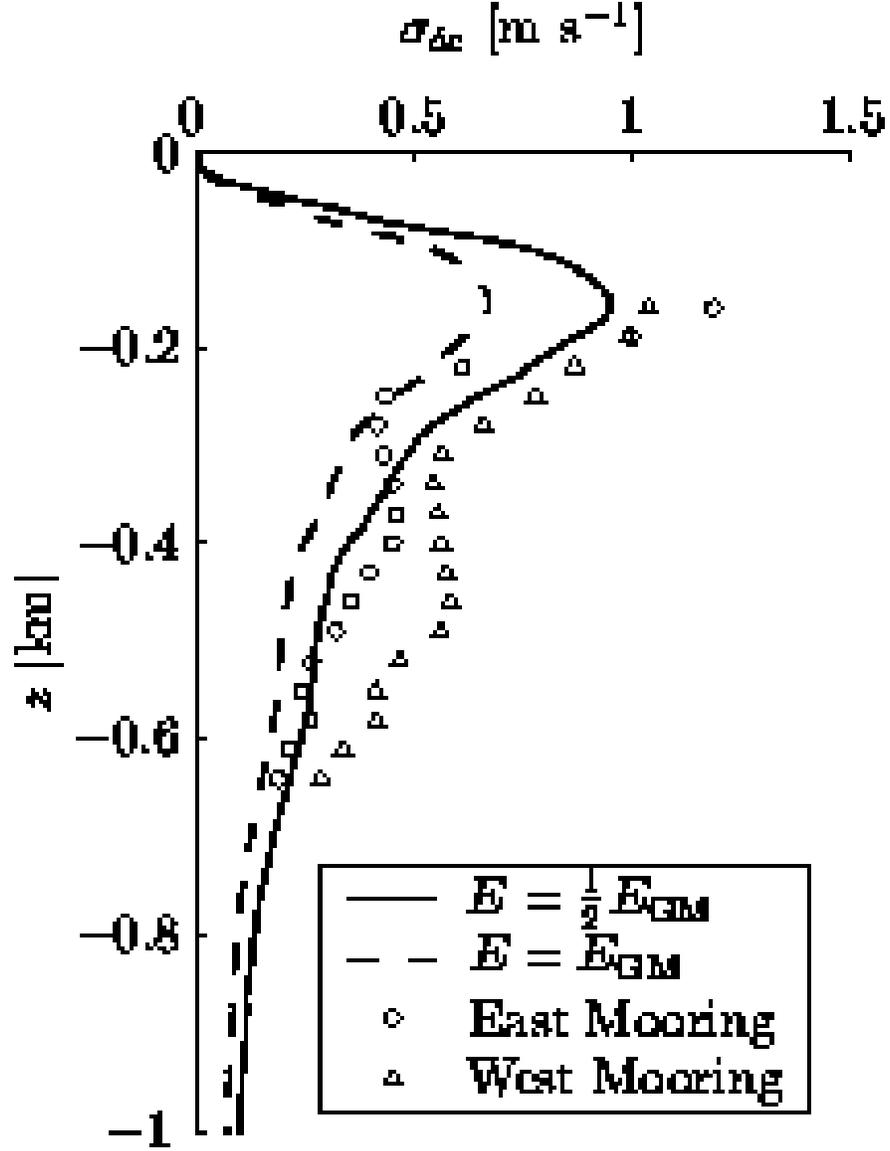}
\caption{
Measured and simulated rms sound speed fluctuations in the AET
environment.  The measurements shown were made near the source (labelled east)
and near the receiving array (labelled west).  The simulated fluctuations are
due to internal waves using both $E = 0.5 \: E_{GM}$ and $E = 1.0 \: E_{GM}$, as
described in the text.
}
\end{figure*}
%%%%%%%%%%%%%%%%%%%%%%%%%%%%%%%%%%%%%%%%%%%%%%%% 
\newpage
%%%%%%%%%%%%%%%%%%%%%%%%%%%%%%%%%%%%%%%%%%%%%%%% 
\begin{figure*}
\includegraphics[width=6.0 in]{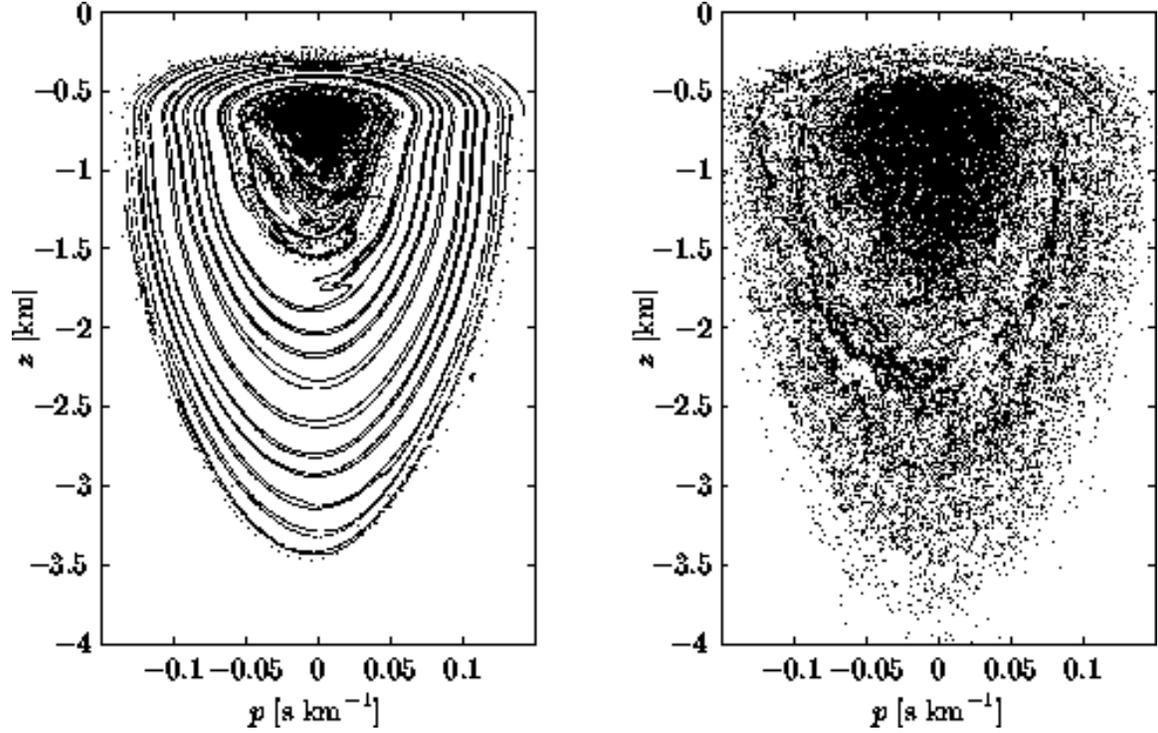}
\caption{
Simulated Lagrangian manifolds in the AET environment without (left
panel) and with (right panel) internal waves.  The ray density used in the plot
on the right and for small $|p|$ in the plot on the left is too sparse to
resolve what should, in each plot, be a smooth unbroken curve that does not
intersect itself.
}
\end{figure*}
%%%%%%%%%%%%%%%%%%%%%%%%%%%%%%%%%%%%%%%%%%%%%%%% 
\newpage
%%%%%%%%%%%%%%%%%%%%%%%%%%%%%%%%%%%%%%%%%%%%%%%% 
\begin{figure*}
\includegraphics[width=5.0 in]{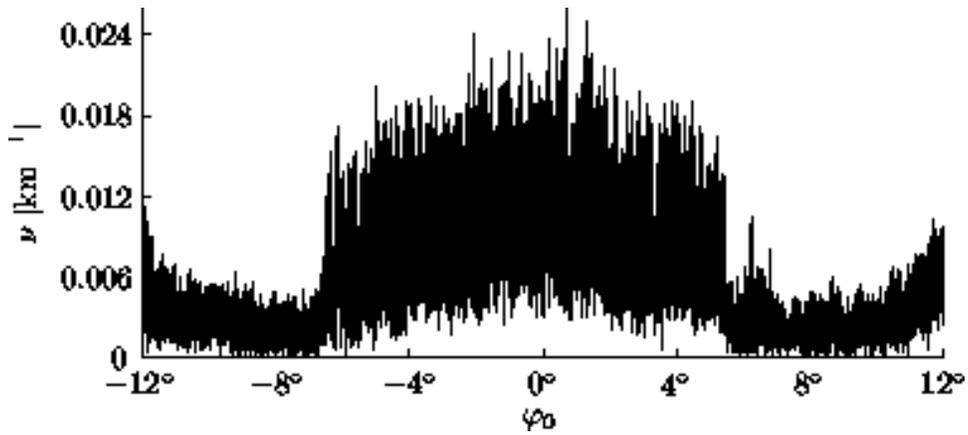}
\caption{
Stability exponents (finite range estimates of Lyapunov exponents)
as a function of ray launch angle in the AET environment.
}
%%%%%%%%%%%%%%%%%%%%%%%%%%%%%%%%%%%%%%%%%%%%%%%% 
\end{figure*}
%%%%%%%%%%%%%%%%%%%%%%%%%%%%%%%%%%%%%%%%%%%%%%%% 
\newpage
%%%%%%%%%%%%%%%%%%%%%%%%%%%%%%%%%%%%%%%%%%%%%%%% 
\begin{figure*}
\includegraphics[height=6.0 in]{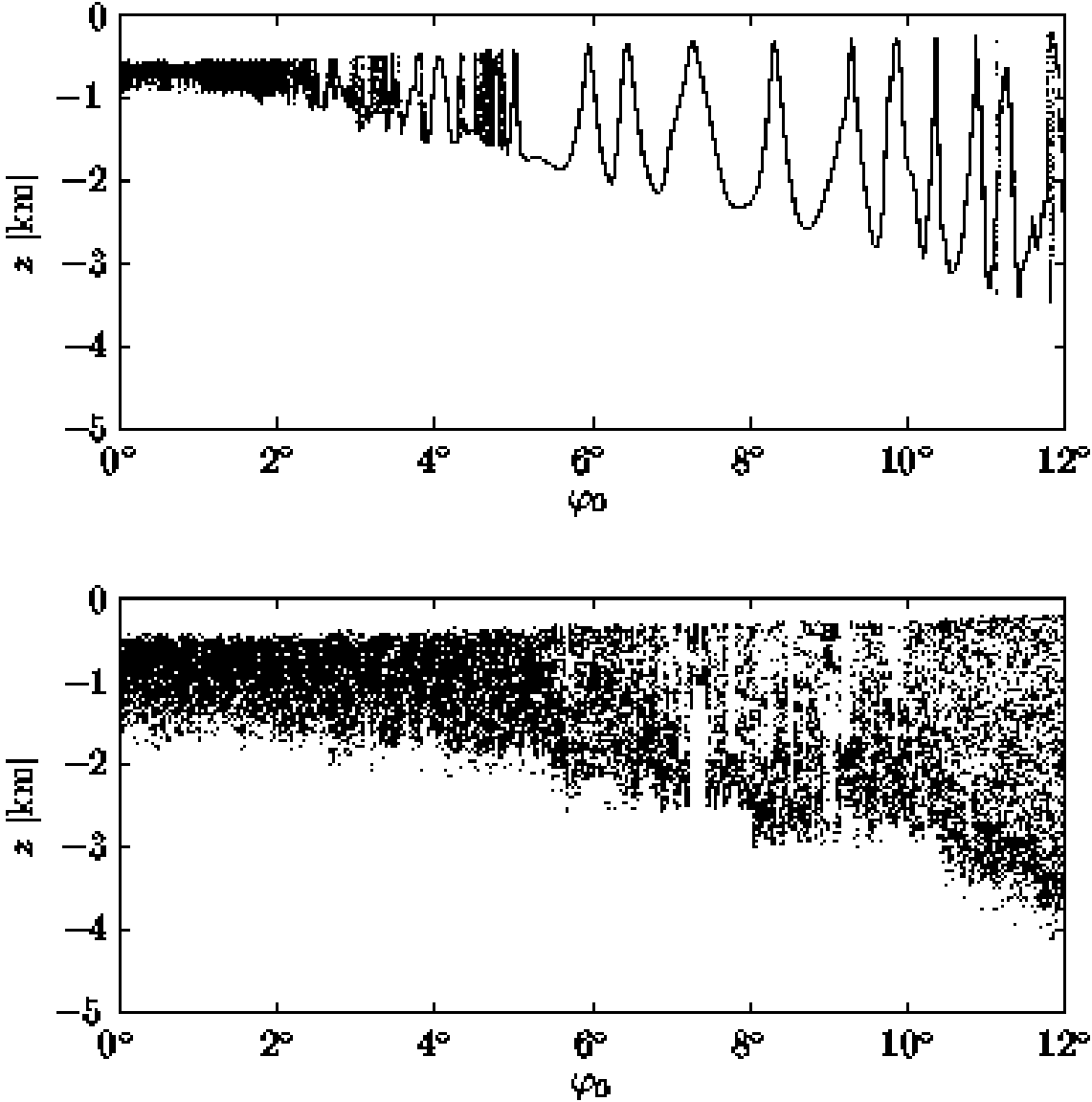}
\caption{
Ray depth vs. launch angle in the AET environment without (upper panel)
and with (lower panel) internal waves.  In both panels
$\Delta \varphi_o = 0.0005^o$.
}
%%%%%%%%%%%%%%%%%%%%%%%%%%%%%%%%%%%%%%%%%%%%%%%% 
\end{figure*}
\newpage
%%%%%%%%%%%%%%%%%%%%%%%%%%%%%%%%%%%%%%%%%%%%%%%% 
\begin{figure*}
\includegraphics[width=6.0 in]{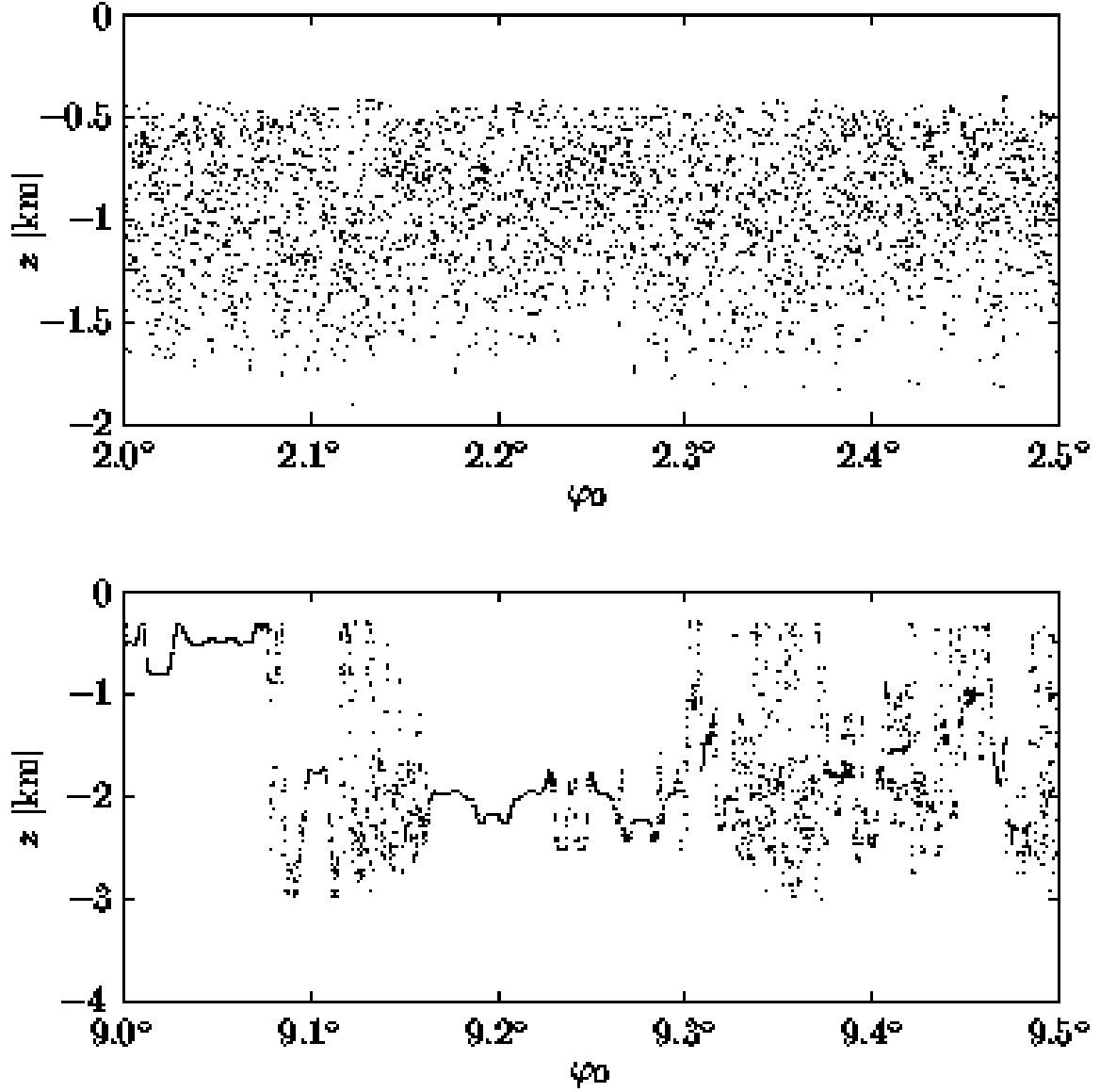}
\caption{
Ray depth vs. launch angle in two small angular bands in the AET
environment with internal waves.  In both panels
$\Delta \varphi_o = 0.000125^o$.
}
\end{figure*}
\newpage
%%%%%%%%%%%%%%%%%%%%%%%%%%%%%%%%%%%%%%%%%%%%%%%% 
\begin{figure*}
\includegraphics[width=6.0 in]{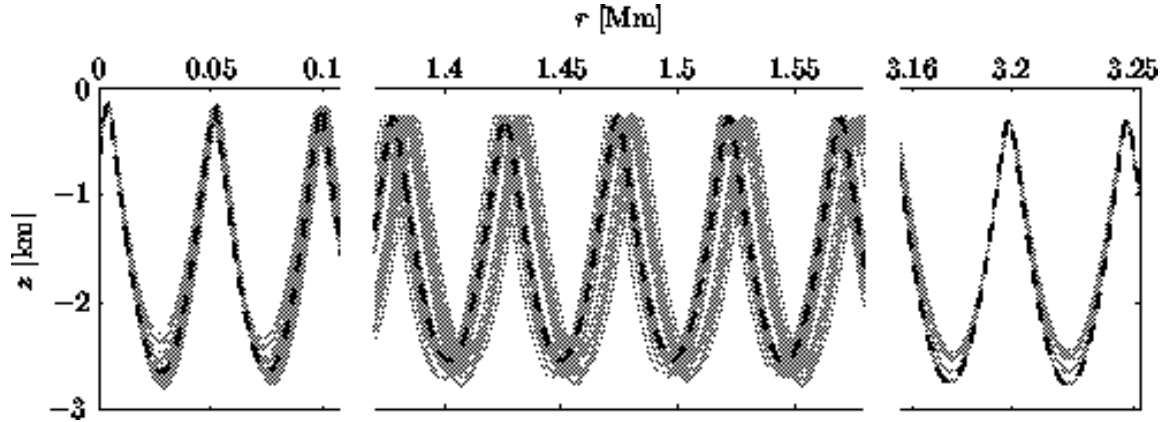}
\caption{
Three segments of the range-depth plane along the AET transmission
path showing a subset of a set of micromultipaths.  The three rays that have
the strongest amplitudes are shown using dashed lines; these rays are not
resolved on this plot.
}
\end{figure*}
\newpage
%%%%%%%%%%%%%%%%%%%%%%%%%%%%%%%%%%%%%%%%%%%%%%%% 
\begin{figure*}
\includegraphics[width=6.0 in]{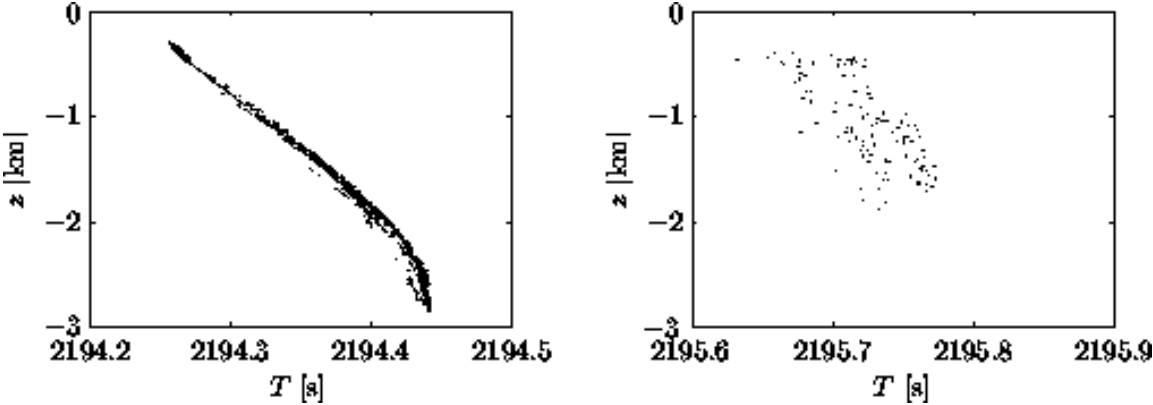}
\caption{
Ray depth vs. travel time in the AET environment with internal waves,
but only for rays with identifiers $+137$ (left panel) and $+151$ (right panel).
The points plotted are a subset of those plotted in the lower panel of Fig. 1.
}
\end{figure*}
\clearpage
\begin{figure*}
\includegraphics[height=2.0 in]{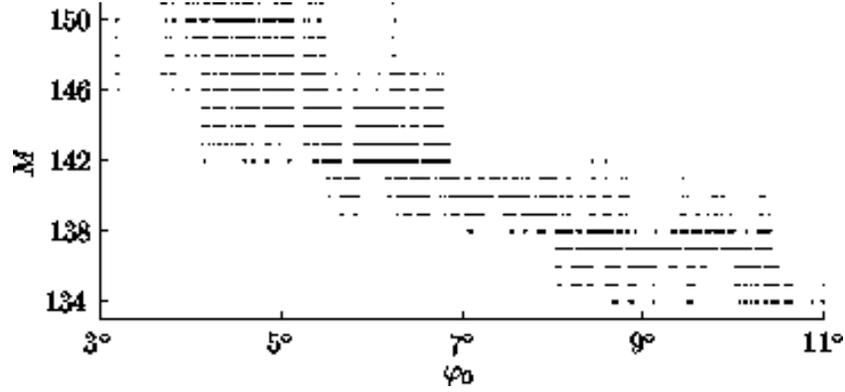}
\caption{
Ray identifier vs. launch angle for a subset of the rays that
are plotted in Fig. 6.
}
\end{figure*}

%%%%%%%%%%%%%%%%%%%%%%%%%%%%%%%%%%%%%%%%%%%%%%%% 
\begin{figure*}
\includegraphics[height=4.0 in]{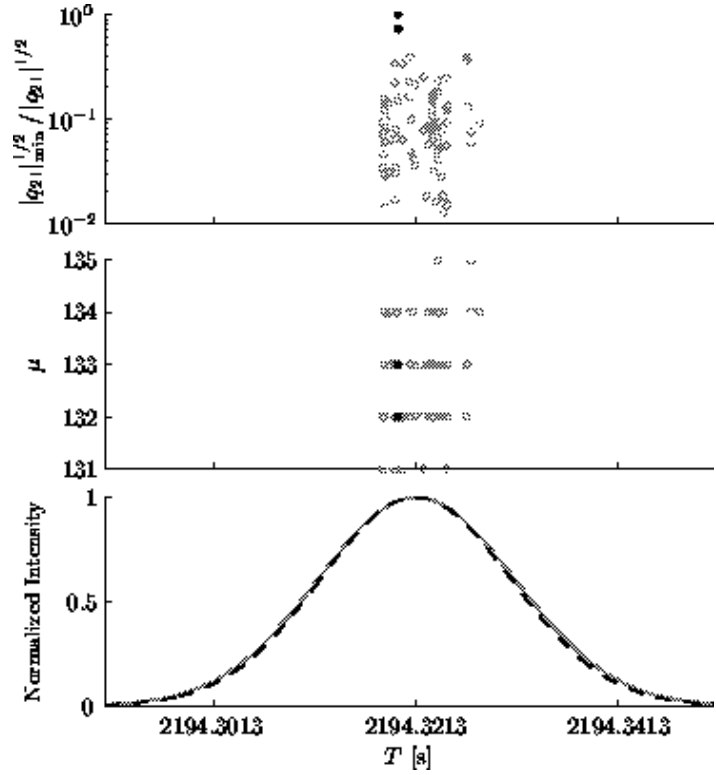}
\caption{
Upper panel: normalized ray intensity vs. travel time for a subset of
the eigenrays (micromultipaths) with identifier +137 that connect the AET
source and a receiver at depth 1005 m at the AET range.  Middle panel: the
corresponding Maslov indices vs. travel time.  Lower panel: envelope of the
waveform synthesized by coherently adding the contributions from all the
micromultipaths shown (solid curve), and the normalized envelope of the AET
source waveform (as seen in the far field; dashed line).  The three rays
whose intensity is largest are shown using solid circles in the upper and
middle panels; the remaining rays are shown using open circles. 
}
\end{figure*}
\newpage
%%%%%%%%%%%%%%%%%%%%%%%%%%%%%%%%%%%%%%%%%%%%%%%% 
\begin{figure*}
\includegraphics[width=4.5 in]{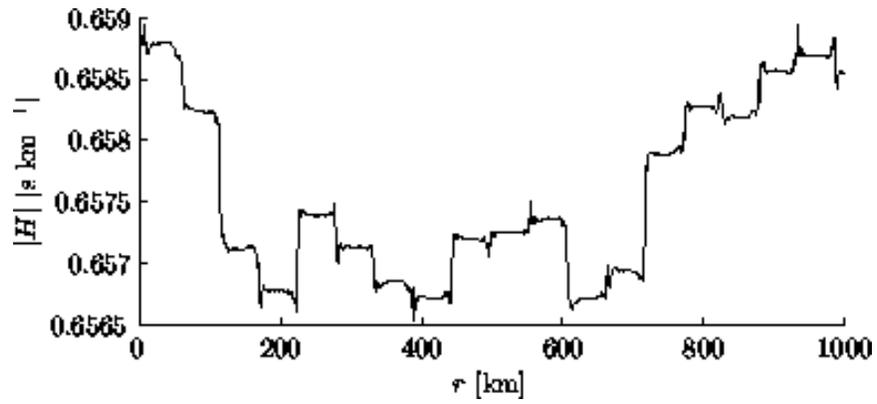}
\caption{
$|H|$ vs. range following a ray emitted on the sound channel axis
with a launch angle of $11^o$ in a canonical environment with an
internal-wave-induced sound speed perturbation superimposed.
}
\end{figure*}
%\newpage
%%%%%%%%%%%%%%%%%%%%%%%%%%%%%%%%%%%%%%%%%%%%%%%% 
\begin{figure*}
\includegraphics[width=3.8 in]{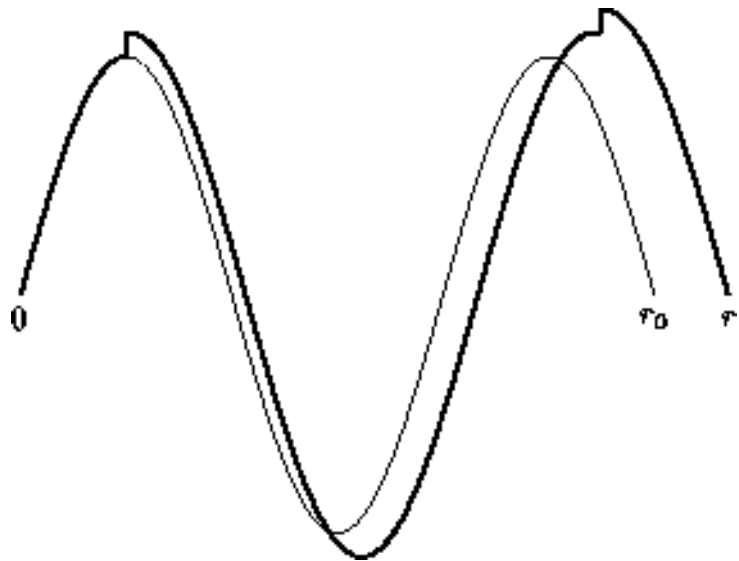}
\caption{
Schematic diagram showing a twice apex-scattered ray (heavy line) and
the path that the same ray would have followed in the absence of apex
scattering events (light line).
}
\end{figure*}
\newpage
%%%%%%%%%%%%%%%%%%%%%%%%%%%%%%%%%%%%%%%%%%%%%%%% 
\begin{figure*}
\includegraphics[width=6.0 in]{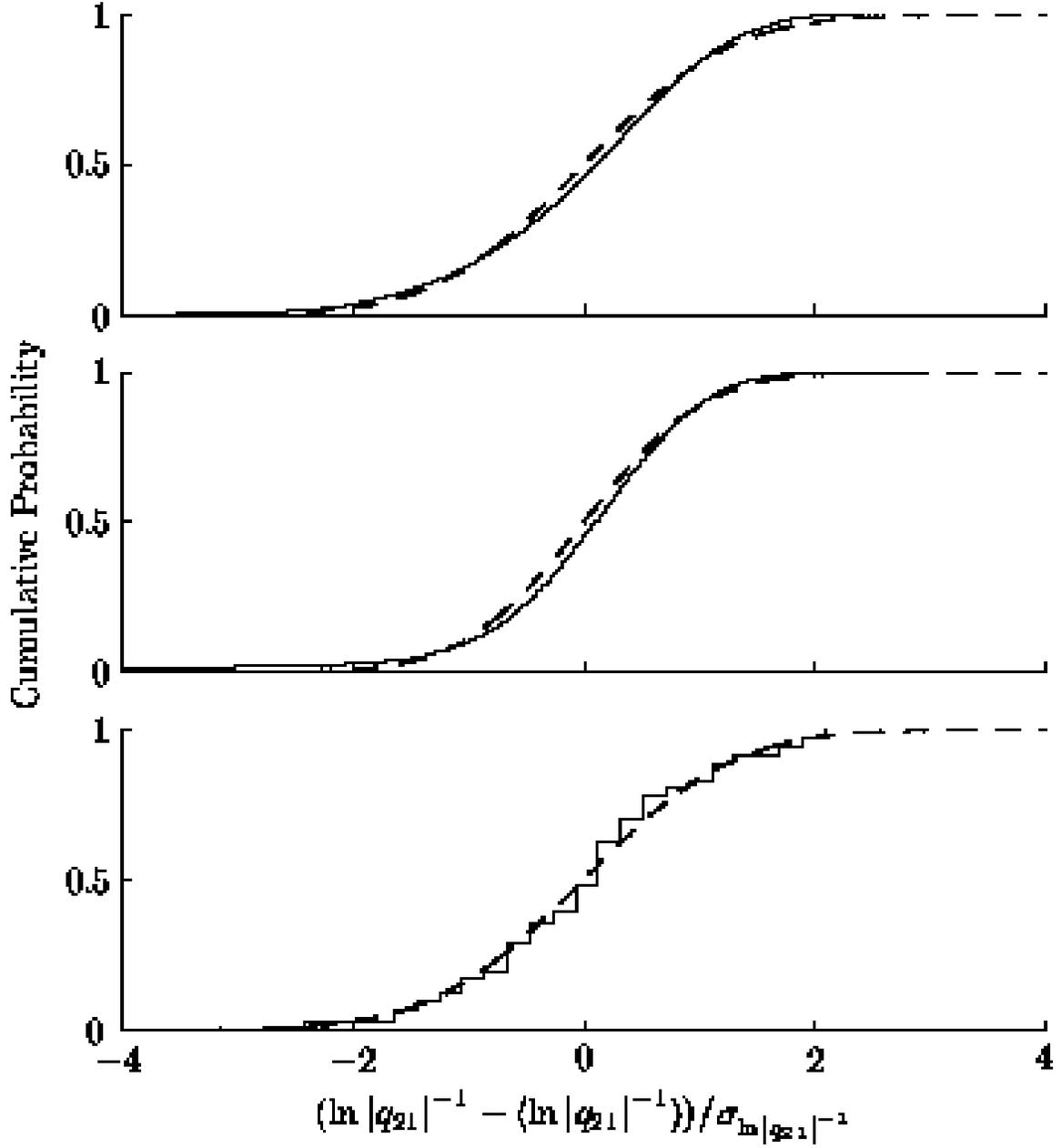}
\caption{
Cumulative probability density of ray intensity in the AET
environment.  Upper panel: flat rays ($|\varphi| \leq 5^o$) sampled uniformly
in launch angle.  Middle panel: steep rays ($6^o \leq |\varphi| \leq 11^o$)
sampled uniformly in launch angle.  Lower panel: eigenrays with $+137$
identifier connecting the AET source and a receiver at depth 1005 m at the
AET range.  In all three panels the dashed lines correspond to lognormal
distributions whose mean and variance approximately match those of the
simulations.
}
\end{figure*}
\newpage
%%%%%%%%%%%%%%%%%%%%%%%%%%%%%%%%%%%%%%%%%%%%%%%% 
\begin{figure*}
\includegraphics[height=4.0 in]{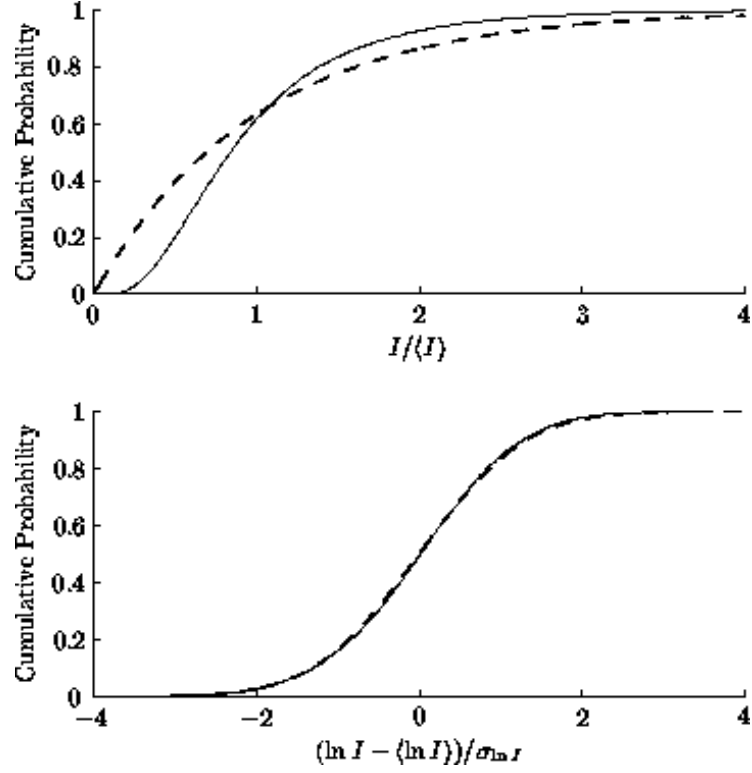}
\caption{
Cumulative probability density of peak wavefield intensity
simulated using a simple model as described in the text.  The same simulated
cumulative density is compared to exponential (upper panel dashed line) and
lognormal (lower panel dashed line) cumulative density functions.  This
figure should be compared to Fig. 14 in Ref.~\cite{AET2}.
}
\end{figure*}
%\newpage
%%%%%%%%%%%%%%%%%%%%%%%%%%%%%%%%%%%%%%%%%%%%%%%% 
\begin{figure*}
\includegraphics[height=5.0 in, angle=-90]{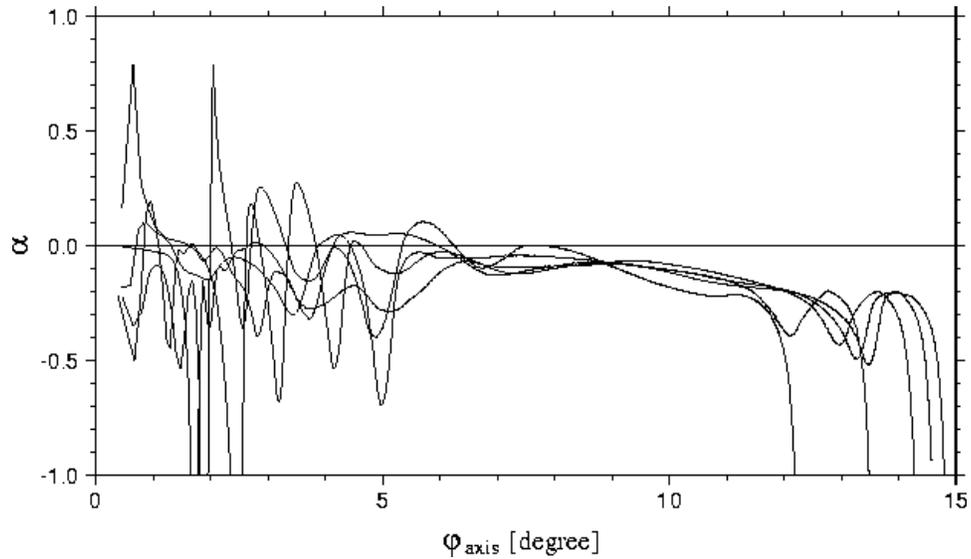}
\caption{
Stability parameter $\alpha$ vs. axial ray angle in the
range-averaged AET environment (dashed curve) and in five 650 km block
range-averages of the AET environment (solid curves).
}
\end{figure*}

\end{document}